\documentclass[showpacs,amsmath,amssymb,twocolumn,pra,longbibliography,superscriptaddress]{revtex4-1}
\usepackage{amssymb}
\usepackage[dvips]{graphicx}
\usepackage{enumerate}
\usepackage{epsfig}
\usepackage{subfigure}
\usepackage{xcolor}
\usepackage[T1]{fontenc}
\usepackage{fullpage}
\usepackage{amsthm,amsfonts,amssymb,amscd,mathrsfs,xspace,framed}
\usepackage{mathrsfs,amsmath}
\usepackage{color}
\usepackage{setspace}
\usepackage{url}
\usepackage{wrapfig}
\usepackage{tikz}
\usepackage{enumitem}
\usepackage{bm}
\usepackage{comment}
\usepackage{txfonts}
\usepackage{array}
\usepackage{color}
\usepackage{braket}

\newcommand*{\field}[1]{\mathbb{#1}}

\begin{document}

\title{Quantum-Computing Architecture based on Large-Scale Multi-Dimensional Continuous-Variable Cluster States in a Scalable Photonic Platform}

\author{Bo-Han Wu$^\dag$}
\email{gowubohan@email.arizona.edu \\ $^\dag$ Equal contributions}

\affiliation{
Department of Physics, University of Arizona, Tucson, Arizona 85721, USA
}

\author{Rafael N. Alexander$^\dag$}
\affiliation{
Center for Quantum Information and Control, University of New Mexico, MSC07-4220, Albuquerque, New Mexico 87131-0001, USA
}

\author{Shuai Liu}
 \affiliation{
Department of Materials Science and Engineering, University of Arizona, Tucson, Arizona 85721, USA
}

\author{Zheshen Zhang}
\affiliation{
Department of Materials Science and Engineering, University of Arizona, Tucson, Arizona 85721, USA
}
\affiliation{
J. C. Wyant College of Optical Sciences, University of Arizona, Tucson, Arizona 85721, USA
}

\date{\today}

\begin{abstract}
Quantum computing is a disruptive paradigm widely believed to be capable of solving classically intractable problems. However, the route toward full-scale quantum computers is obstructed by immense challenges associated with the scalability of the platform, the connectivity of qubits, and the required fidelity of various components. One-way quantum computing is an appealing approach that shifts the burden from high-fidelity quantum gates and quantum memories to the generation of high-quality entangled resource states and high fidelity measurements. Cluster states are an important ingredient for one-way quantum computing, and a compact, portable, and mass producible platform for large-scale cluster states will be essential for the widespread deployment of one-way quantum computing. Here, we bridge two distinct fields---Kerr microcombs and continuous-variable (CV) quantum information---to formulate a one-way quantum computing architecture based on programmable large-scale CV cluster states. The architecture can accommodate hundreds of simultaneously addressable entangled optical modes multiplexed in the frequency domain and an unlimited number of sequentially addressable entangled optical modes in time domain. One-dimensional, two-dimensional, and three-dimensional CV cluster states can be deterministically produced. We note cluster states of at least three dimensions are required for fault-tolerant one-way quantum computing with known error-correction strategies. This architecture can be readily implemented with silicon photonics, opening a promising avenue for quantum computing at a large scale.

\end{abstract}

\maketitle

\section{Introduction}

Quantum computing is deemed a disruptive paradigm for solving many classically intractable problems such as factoring big numbers~\cite{Shor94}, data fitting~\cite{Wiebe12}, combinatorial optimization~\cite{Djidjev18}, and boson sampling~\cite{Aaronson10}. The development of quantum-computing platforms has significantly progressed over the last decade~\cite{Ganzhorn19,Monroe95,Wallraff04,Zhou17,Loss98}, but outstanding challenges associated with the system scalability, fidelity of quantum gates, and controllability of qubits remain. To date, there has not been a single quantum-computing platform that successfully addresses all these challenges.

One-way quantum computing~\cite{Raussendorf01} is an intriguing approach to obviate the demanding requirement on quantum-gate fidelity~\cite{Browne05}. Unlike quantum-computing schemes based on quantum gates, the quantum logic in one-way quantum computing is implemented via measuring a highly entangled state known as a \emph{cluster state}~\cite{Briegel01}. Measurements are implemented sequentially, so that the measurement basis at a given step may be adaptively chosen based on outcomes of prior measurements. Thus, given access to high quality entangled resource states and high-fidelity measurements, the need for active quantum gates is eliminated.

One-way quantum computing can be implemented in different platforms, and is particularly well suited to quantum-photonic architectures because: first, photons are robust quantum-information carriers even at room temperature; second, quantum measurements on photons are well developed---they can be precisely controlled and efficiently read out; and third, photons can be readily transmitted over long distances to link distributed quantum-computing and quantum-sensing devices without requiring extra quantum-information transductions. A barrier to photonic one-way quantum computing, however, lies in the generation of large-scale, high-quality cluster states. Photonic one-way quantum computing based on DV cluster states, typically based on dual-rail encoding on single photons, has been theoretically studied~\cite{Nielsen04,Raussendorf01,Joo07} and verified in proof-of-concept experiments~\cite{Walther05,Kiesel05,Biggerstaff09,Vallone07}. Scaling up the size of DV cluster states, however, is impeded by a lack of deterministic means for their generation. A mainstream mechanism to produce DV cluster states based on spontaneous parametric down-conversion (SPDC) in nonlinear crystals followed by non-deterministic post-selection suffers from an exponentially small state-generation success rate as the size of the DV cluster state increases. Deterministic means of generating large-scale DV cluster states from single quantum emitters have only appeared in recent theoretical works~\cite{Russo19,Buterakos17}. 

CV states are encoded into continuous quadratures of bosonic modes. Like DV systems, superdense coding~\cite{Braunstein00}, quantum teleportation~\cite{Braunstein98}, and quantum cryptography~\cite{Ralph99} have been demonstrated in CV systems. Moreover, one-way quantum computing can also be generalized to CVs~\cite{Menicucci06}. An appealing feature of this approach is that large-scale entangled states can be deterministically generated at a large scale. Indeed, CV-cluster-state sources have been studied in the frequency domain~\cite{Menicucci08,Chen14} and the time domain~\cite{Menicucci11,Alexander18,Sabapathy18,Yokoyama13,Yoshikawa16,Asavanant19}. A recent experiment of frequency-multiplexed CV cluster states demonstrated 60 {\em simultaneously} accessible spectral modes~\cite{Chen14}. In the time domain, temporal modes can be addressed {\em sequentially}, enabling demonstrations of cluster states made of 10,000 modes~\cite{Yokoyama13} and over one-million modes~\cite{Yoshikawa16}. Though large in scale, the aforementioned demonstrations all generated one-dimensional cluster states, which are insufficient for universal one-way quantum computing. More recently, two-dimensional time-multiplexed CV cluster states were generated by Asavanant {\em et al.}~\cite{Asavanant19} and Larsen {\em et al.}~\cite{Larsen19}. The utility of such 2D CV cluster states in one-way quantum computing is, however, constrained by the shorter of the two dimensions. Extending this dimension comes at the price of potentially introducing additional losses, limiting the potential scalability of time-multiplexing in more than one dimension. 
Hybrid time-frequency multiplexed CV cluster states~\cite{Alexander16, Humphreys14} would significantly enlarge the size of the shorter dimension, but obtaining phase references to simultaneously access all spectral modes remains an outstanding open problem.

A key factor in assessing the feasibility of fault-tolerant measurement-based quantum computation is the amount of squeezing available in a CV cluster state~\cite{Gu09, Alexander14}. The amount of required squeezing depends on the form of error correction used~\cite{Menicucci14}. Recent work has highlighted the possibility of using a combination of robust bosonic qubits, known as the \emph{Gottesman-Kitaev-Preskill} (GKP) encoded qubits~\cite{Gottesman01}, and 3D entangled structures to implement fault-tolerant quantum computation~\cite{Fukui17, Vuillot19, Fukui19, Noh19}. While the former have recently been demonstrated experimentally~\cite{Touzard19, Fluhmann19}, the latter still presents a challenge. As such, a platform that generates 3D CV cluster states would be an enabler for fault-tolerant quantum computing. 

In this article, we bridge two distinct fields, Kerr-soliton microcombs and CV quantum information, to formulate a one-way quantum-computing architecture based on large-scale 3D CV cluster states generated in a scalable quantum-photonic platform. In the proposed architecture, third-order ($\chi^{(3)}$) Kerr nonlinearity is utilized with both time and frequency multiplexing to produce reconfigurable 1D, 2D, or 3D CV cluster states. Frequency multiplexing can provide access to hundreds of simultaneously accessible, highly-connected spectral modes, whereas the time multiplexing allows for sequential access to an unlimited number of temporal modes. By virtue of large bandwidth ($\sim$GHz) of the spectral modes, the quantum-photonic platform offers the scalability and robustness required to produce large-scale 3D CV cluster states for fault-tolerant quantum computing. A unique advantage of our approach---which uses $\chi^{(3)}$ Kerr nonlinearity---is that we can generate a frequency-comb soliton suitable for acting as a local phase reference for all spectral modes, thereby solving a key challenge that faced previous work on frequency-multiplexed CV cluster states. Access to a large number of spectral modes enables us to reach a scale required to see a truly 3D structure, without introducing prohibitively high loss.

In Sec.~\ref{SEC:Architecture}, we will first describe the mechanism to obtain classical frequency-comb phase references followed by elaborating the architecture for generating programmable CV cluster states. We proceed in Sec.~\ref{SEC:ArchAn} to analyze device parameters that determine the scale and quality of the generated CV cluster states. The universal one-way quantum computing framework tailored to the 3D CV cluster state is formulated in Sec.~\ref{SEC:QC}.

\section{The architecture} \label{sec:arch}
The microring resonator (MR) is the workhorse of our photonic platform, providing the means of generating both a classical frequency-comb reference and large-scale CV cluster states. In this section, we discuss the physics relevant to utilizing MRs for the generation of both classical and non-classical states of light.
\label{SEC:Architecture}

\subsection{Classical frequency-comb phase references}
\label{SEC:Reference}
\begin{figure}[h]
	{\centering\includegraphics[width=0.85\linewidth]{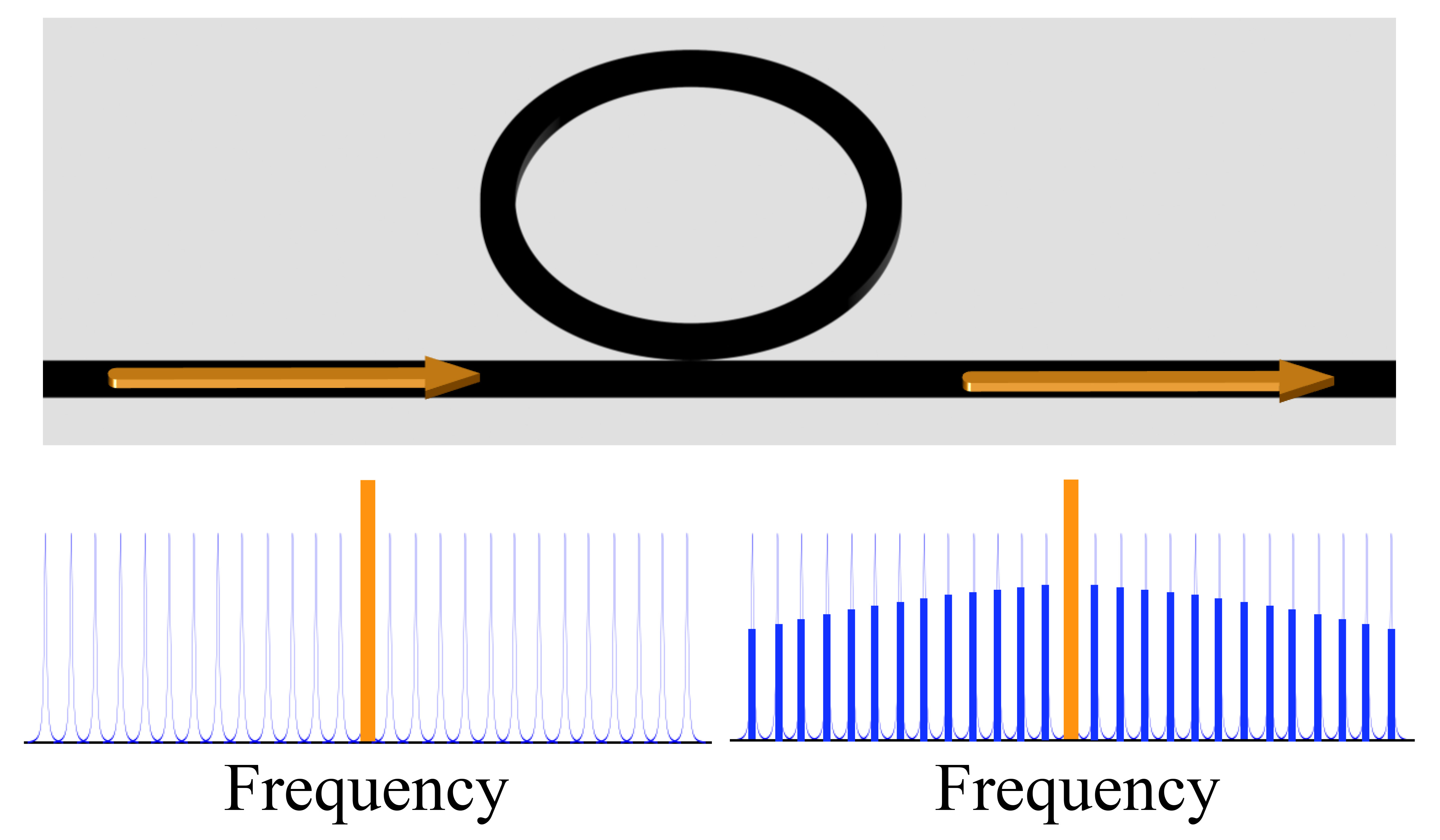}\\}
	\caption{\label{FIG:Comb} (Top panel) Classical frequency-comb generation. (Bottom panel) The spectra for the input pump (orange bar) and the output side-mode fields (blue bars).
	}
\end{figure}

A continuous-wave (c.w.) pump field is sent through a bus waveguide and coupled into the MR, as shown in Fig.~\ref{FIG:Comb}. The power of the in-coupled field is greatly enhanced by an appropriate quality ($Q$) factor of the MR.  Above the parametric oscillation threshold, side-mode fields (represented as blue bars in Fig.~\ref{FIG:Comb}) are created via four-wave mixing (FWM). The generated side-mode fields then couple with the pump field to create more side-mode fields via stimulated FWM. Moreover, provided that the power of the generated side-mode fields are above the cavity threshold, they also serve as new pump sources that, in turn, generate other side-mode fields. Ultimately, this cascading FWM process will lead to an extensively-extended spectrum profile, as shown in Fig.~\ref{FIG:Comb}.
Locking the phase of each frequency tooth leads to the generation of a Kerr-soliton~\cite{Lamont13,Coen13}, which can address the corresponding spectral mode of the CV cluster state in coherent quantum measurements.

\subsection{Quantum CV cluster-state sources}
We now describe our method for generating zero-, one-, two-, and three-dimensional CV cluster states. This involves sending a c.w.~pump field, whose power is {\em below} the parametric oscillation threshold, through the configuration described in Sec.~\ref{SEC:Reference}. Choosing an input pump power level below the cavity oscillator threshold brings multiple benefits. First, operating at a lower power level is more energy efficient; second, this reduces thermally-induced instabilities; and third, the mean fields of the quantum modes would otherwise be very large above the oscillation threshold, creating a barrier to quantum-limited homodyne detection. We then describe a programmable photonic platform that can switch between generating a variety of different CV cluster states with different dimensions simply by tuning the phase of various MZIs. In our scheme, we pump the MRs at even spectral modes while detecting the output fields at only odd modes.

Throughout this article, we represent the multimode Gaussian states generated using the graphical notation introduced in Ref.~\cite{Menicucci11} and summarized in Appendix~\ref{sec:graphnot}.

\subsubsection{Entangled spectral mode pairs: 0D CV cluster states}
\label{SEC:ODCluster}
\begin{figure}[h]
	{\centering\includegraphics[width=0.85\linewidth]{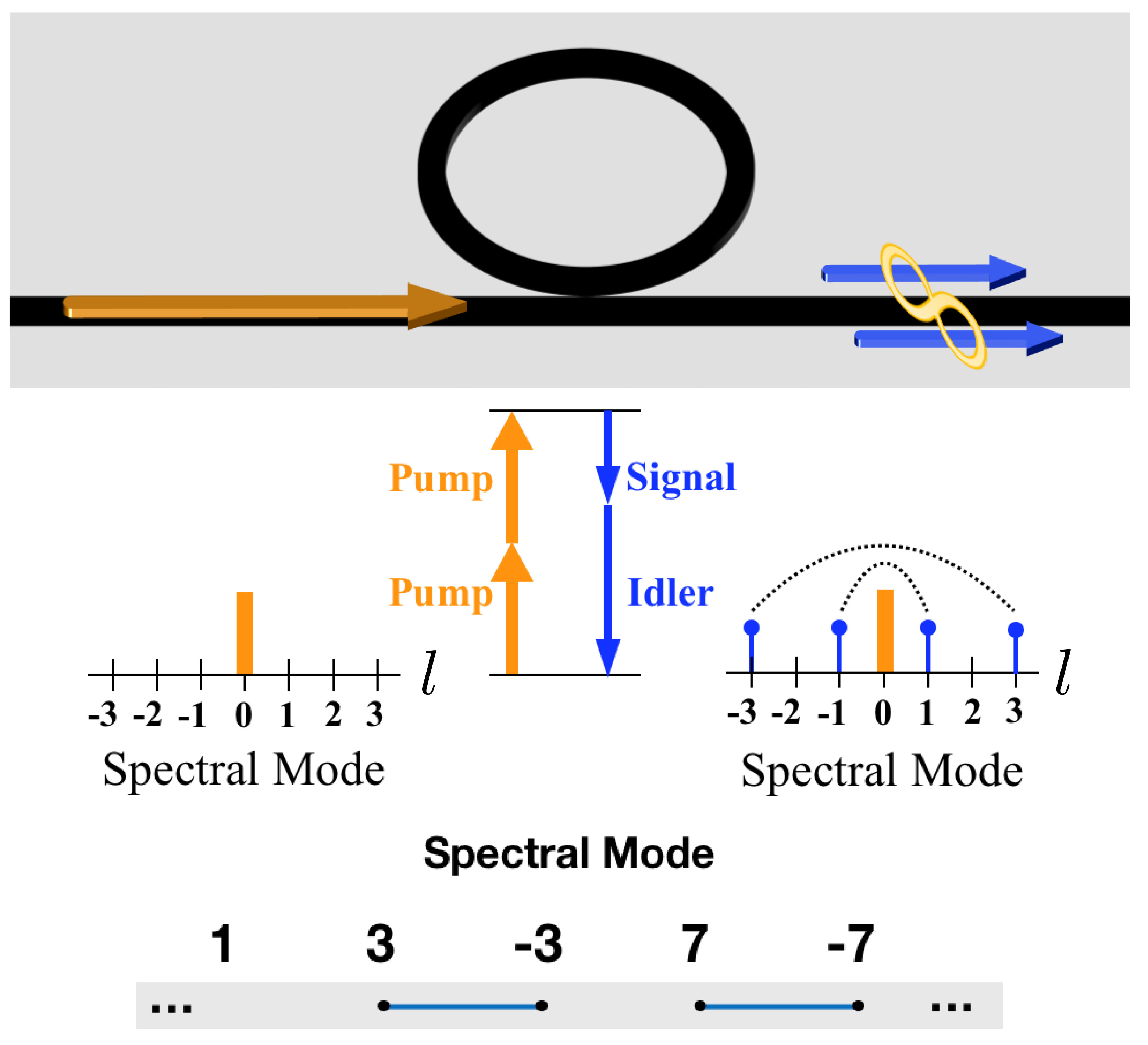}\\}
	\caption{\label{FIG:0DCluster} Generation of the 0D cluster state. Middle panel illustrates the FWM process in the MR. The bottom panel is the graph representation of the output state with $C=1$.
	}
\end{figure}

 The FWM process couples different cavity spectral modes, creating side-mode fields in a pair-wise fashion. As shown in Fig.~\ref{FIG:0DCluster}, pairs of pump photons at spectral mode $l=0$ are converted into signal photons at spectral modes $-l=-1,-3,-5,\cdots$ and idler photons at spectral modes $l=1,3,5,\cdots$. These modes become entangled with each other. More specifically, they become {\em two-mode squeezed states}, which are equivalent to two-mode CV cluster states via application of local phase shifts~\cite{Menicucci11}. 

\subsubsection{1D CV cluster states}
\begin{figure}[h]
	{\centering\includegraphics[width=0.85\linewidth]{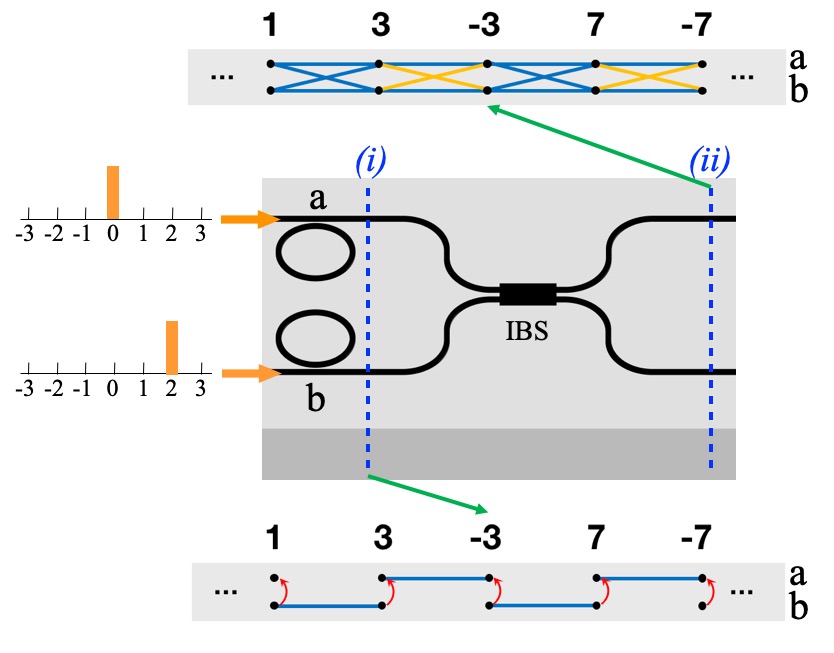}\\}
	\caption{\label{FIG:1DCluster} Generation of the 1D cluster state. Top and bottom inserted panels depict the graph-state representations right after stage $(i)$ and $(ii)$. The upper rail is for spatial mode `a', and the lower rail is for spatial mode `b'. $C= 1$ for stage $(i)$ and $C= 1/2$ for stage $(ii)$.}
\end{figure}

The 1D cluster-state source consists of two identical sets of 0D configurations connected by a 50:50 integrated beamsplitter (IBS). This IBS is designed so that it is capable of coupling the fields across a wide-frequency range~\cite{Zhang13}. The two MRs are pumped at different cavity spectral modes: spatial mode `a' is at $l=0$ and `b' is at $l=2$. The frequency offset of these two pumps results in each frequency being connected by an entangled pair to its neighbor, as shown in stage $(i)$ of Fig.~\ref{FIG:1DCluster}. A one-dimensional entangled CV cluster state is produced in the frequency domain, as shown in stage $(ii)$ of Fig.~\ref{FIG:1DCluster}. This is known as the {\em dual-rail wire} and is a resource for single-mode CV one-way quantum computing~\cite{Alexander14}. States of this type have been successfully generated using bulk-optics setups~\cite{Chen14, Yokoyama13, Yoshikawa16}.

\subsubsection{2D CV cluster states}
By extending the setup from the 1D case by including an additional unbalanced Mach-Zehnder interferometer (UMZI), delay line (DL) and one 50:50 IBS, we are able to generate a 2D universal CV cluster state known as the {\em bilayer square lattice}~\cite{Alexander16}. This approach is equivalent to the bulk-optics scheme proposed in Ref.~\cite{Alexander16} to produce a ({\em frequency})$\times$({\em time}) 2D CV cluster state, but the large MR bandwidth now allows for a shorter delay line (DL) that can be integrated on a photonic chip.
\begin{figure}[h]
    {\centering\includegraphics[width=1\linewidth]{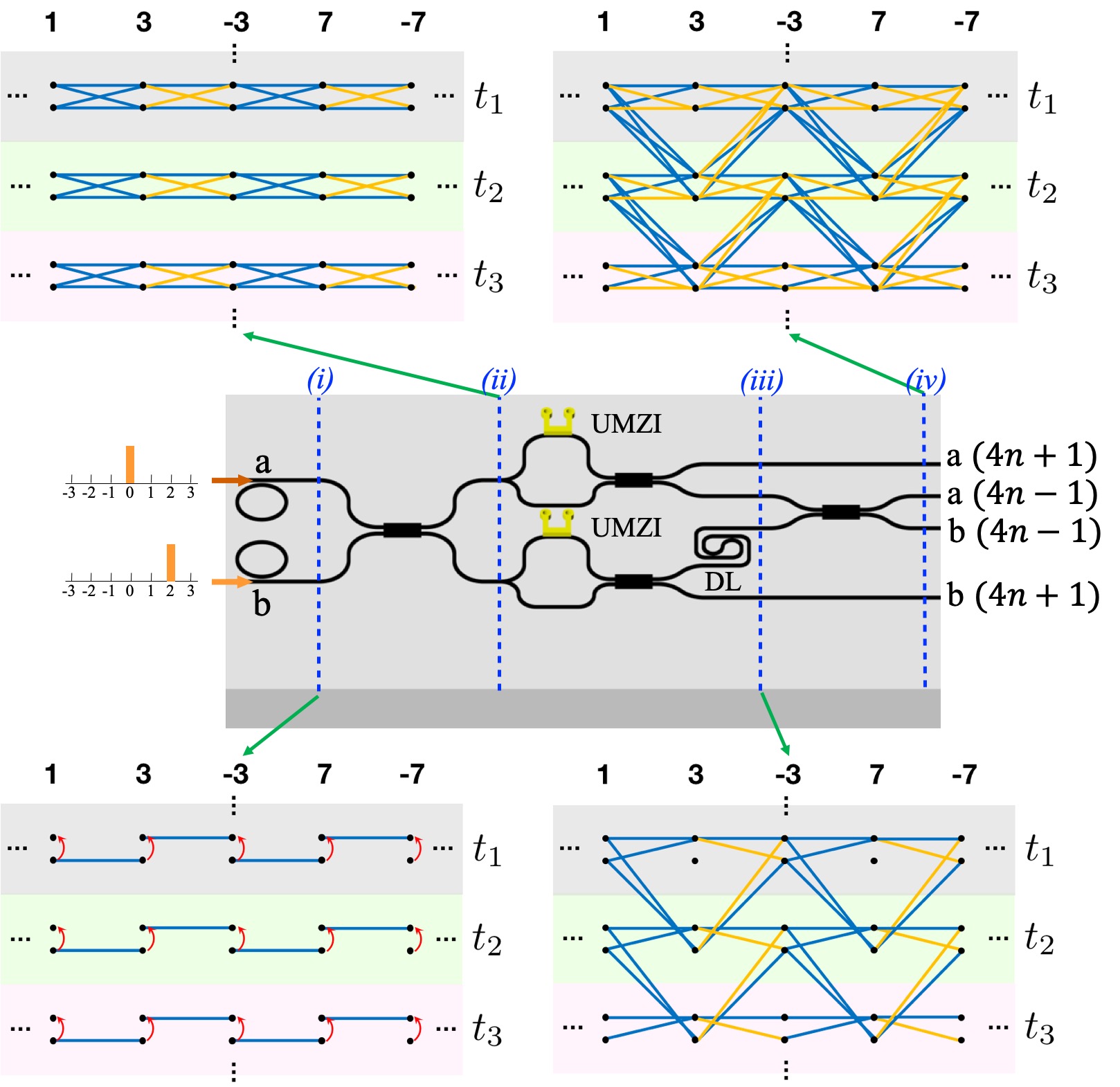}\\}
    \caption{\label{FIG:2DCluster} Generation of the 2D cluster state. Gray, green, and magenta shaded areas in the inserted figures denote the temporal modes $t_{1}, t_{2},t_{3}\in T$, where $t_{2}=t_{1}+\delta t$ and $t_{3}=t_{2}+\delta t$, and $\delta t$ is the time delay due to the DL. For stages $(i-iv)$, $(ii)$, $(iii)$, and $(iv)$, $C = 1$, $1/2$, $1/2$, and $1/2\sqrt{2}$, respectively. The labels a$\,(4n+1)$, a$\,(4n-1)$, b$\,(4n+1)$ and b$\,(4n-1)$ indicate the spatial mode indices (`a' and `b') followed by the spectral mode index, where $n\in \mathbb{Z}$. Electrodes for the UMZI are shown in yellow.}
\end{figure}

Stage $(i)$ and $(ii)$ in Fig.~\ref{FIG:2DCluster} are the same as in Fig.~\ref{FIG:1DCluster}. After stage $(ii)$, the four ports are processed by two UMZIs. The length difference between the two arms of each UMZI is specially designed so that the spectral modes $l=4n+1$ are spatially separated from the spectral modes $l=4n-1$, where $n\in\mathbb{Z}$~\cite{Glockl04,Huntington05,Alexander16}. The two UMZIs are fine tuned using electrodes via the thermal-optical effect~\cite{Elshaari16,Xue16}.

After stage $(ii)$, the field in one arm is temporally delayed. This arrangement extends the spectral entanglement across modes with different temporal indices. The state after stage $(iii)$ is shown in Fig.~\ref{FIG:2DCluster}. In stage $(iv)$, the middle two arms are mixed by another 50:50 IBS, generating a 2D CV cluster state. A closely related CV cluster state was recently generated in the time domain~\cite{Asavanant19} using a long DL.

\subsubsection{3D CV cluster states} \label{sec:3dCVCSgen}
\begin{figure*}
	{\centering\includegraphics[width=1\linewidth]{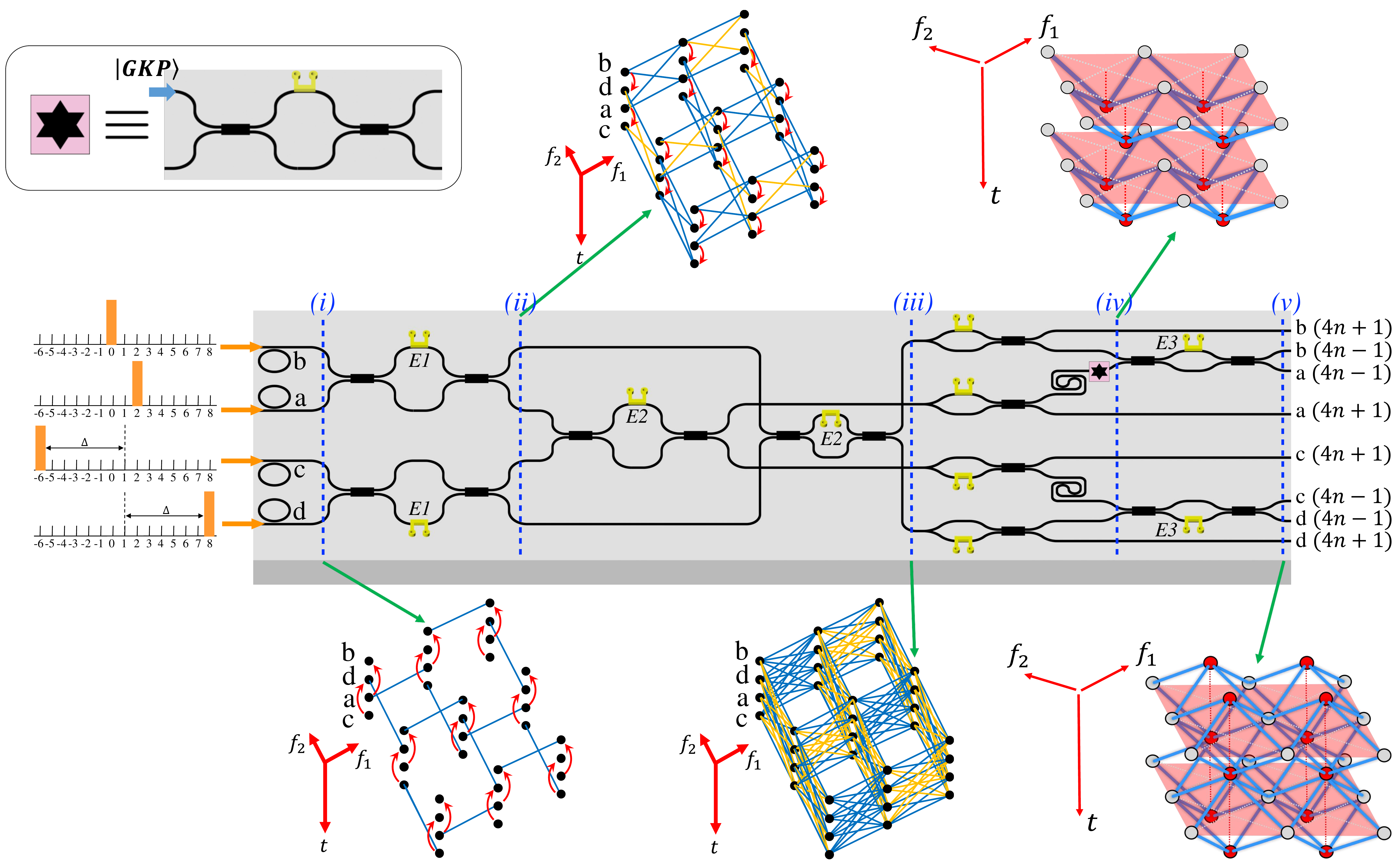}\\}
	\caption{\label{FIG:Programmable} Generation of programmable CV cluster states. The inserted figures show the graph states at stages $(i-v)$. The graphs at stages $(iv)$ and $(v)$ are \emph{coarse-grained} so that each node contains four modes and each edge represents many connections between two particular macronodes (group of four modes). The magenta box indicates where single-mode input states, such as GKP qubit ancilla states, can be injected into the 3D lattice via the BMZI. For stages $(i-v)$, $\mathcal{C} = 1, 1/2, 1/4, 1/4$ and $(4\sqrt{2})^{-1}$, respectively.}
\end{figure*}
\begin{figure*}
	{\centering\includegraphics[width=0.9\linewidth]{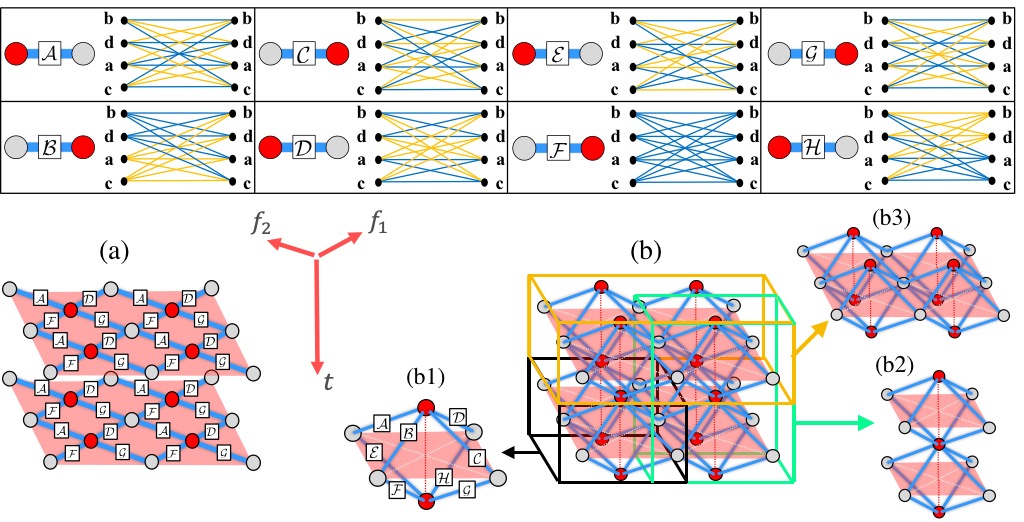}\\}
	\caption{\label{FIG:3DCVcsgraph2} Introduction of the macronodes for Fig.~\ref{FIG:Programmable}. Each edge is labelled corresponding to the legend shown above. (a) The single sheet is equivalent to the graph state at stage $(iii)$, where multiple copies of the quad rail lattice cluster state multiplexed in time. $\mathcal{C}= 1/4$ (b) 3D cluster state at stage $(v)$. (b1) Unit cell of the 3D cluster state. (b2) Single space-like slice of the 3D cluster state. (b3) Single time-like slice of the 3D cluster state. $\mathcal{C} = (4\sqrt{2})^{-1}$.}
\end{figure*}

A 3D CV cluster state can be generated using the setup shown in Fig.~\ref{FIG:Programmable}. Part of the setup consists of two copies of the 2D cluster state setup, but all the 50:50 IBSs are replaced by balanced Mach-Zehnder interferometers (BMZIs). The Mach-Zehnder interferometers are tuned to act as 50:50 IBSs. This replacement will be relevant in the next section where we discuss how to tune the Mach-Zehnder inteferometers in order to make CV cluster states with the same chip. These two copies are coupled together with two additional BMZIs at stage $(ii)$ in Fig.~\ref{FIG:Programmable}.

Spatial modes $\{\text{a},\text{b},\text{c},\text{d}\}$ are pumped at spectral modes $l=0, 2, 1+\Delta, 1-\Delta$, respectively, where $\Delta \in \{2n+1, n\in\field{N}\}$ is a free parameter that sets the length of one lattice direction in frequency, as described below. At stage $(i)$, the state consists of a collection of entangled pairs. At stage $(ii)$, each mode has passed through a BMZI, resulting in a collection of dual-rail wire graphs, just like in the 1D case. At stage $(iii)$, two additional BMZIs stitch these wires together to create a 2D square lattice embedded on a cylinder with circumference $\Delta$  and length set by the overall bandwidth of the experiment. This state is known as the quad-rail lattice~\cite{Menicucci11, Wang14}. In fact, the generation circuit until this point is the same as was proposed in Ref.~\cite{Wang14}. At stage $(iv)$, one quarter of the modes are delayed by one time step. This is analogous to the use of DL in 2D case. The result is the 3D CV cluster state shown in Fig.~\ref{FIG:Programmable} and further elaborated in Fig.~\ref{FIG:3DCVcsgraph2}. Finally, two additional 50:50 IBSs are applied on four of the resulting fields. This is a ({\em frequency})$\times$({\em frequency})$\times$({\em time}) lattice. Single mode input states, such as GKP ancilla states, can be injected into the cluster state by an input port indicated on the chip in Fig.~\ref{FIG:Programmable}. Use of such states for universal fault-tolerant quantum computation will be discussed later in Sec.~\ref{sec:uniGKP}.

The 3D structure of the cluster state becomes apparent when the modes are combined into groups of four, referred to as \emph{macronodes}.

\subsection{Programming the 3D CV cluster state chip for other lattices}
\begin{figure*}
	{\centering\includegraphics[width=1\linewidth]{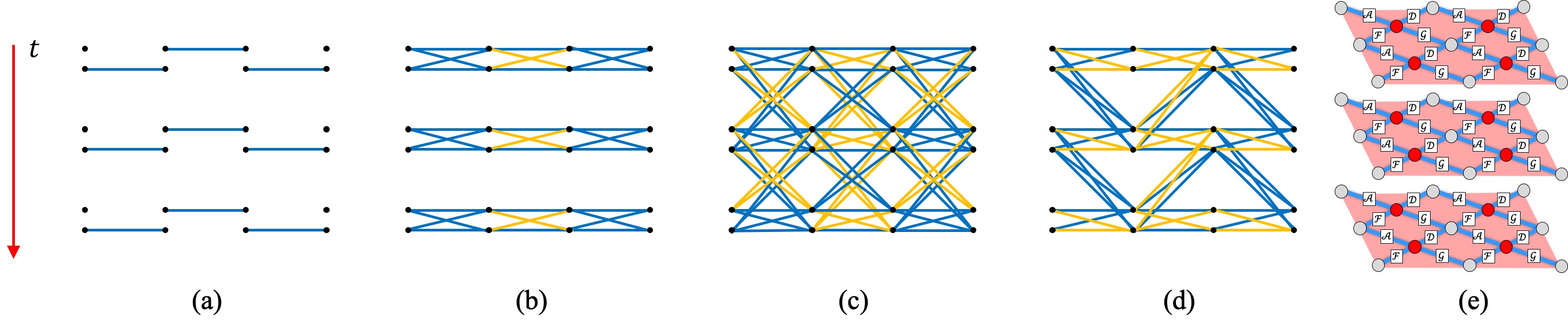}\\}
	\caption{\label{FIG:Reprochip} (a) A (\emph{frequency})$\times$(\emph{time}) array of two-mode CV cluster states. $\mathcal{C}=1$. (b) A train of frequency entangled dual-rail wire cluster states. $\mathcal{C}=1/2$. (c) A (\emph{frequency})$\times$(\emph{time}) entangled 2D resource~\cite{Larsen19}. $\mathcal{C} = 1/4$. (d) The bilayer square lattice cluster state~\cite{Alexander16}. $\mathcal{C}=(2\sqrt{2})^{-1}$. (e) A train of quad-rail lattice cluster states~\cite{Wang14}. $\mathcal{C}=1/4$. }
\end{figure*}
Besides the 3D CV cluster state, the chip proposed in Fig.~\ref{FIG:Programmable} is able to generate CV cluster states of any dimension from 0D to 3D by controlling phase shifts via the electrodes, E1, E2, and E3 in Fig.~\ref{FIG:Programmable}. These electrode control the relative phases between the two arms of BMZIs such that the splitting ratios are tuned.

First, 0D cluster states, i.e., a collection of pairwise entangled states, can be generated by setting the splitting ratios of all BMZIs to be 100:0. These are shown in Fig.~\ref{FIG:Reprochip}(a). 

To make many copies of 1D entangled states in frequency as shown in Fig.~\ref{FIG:Reprochip}(b), one tunes the splitting ratios of the BMZIs at E1, 50:50, and the BMZIs at E2 and E3, 100:0. 

In order to make two copies of 2D cluster states shown in Fig.~\ref{FIG:Reprochip}(c), one sets the splitting ratios of the BMZIs at E2, 100:0 (rather than 50:50) and modifies the BMZIs at E1, and E3, 50:50. These 2D cluster states are ({\em frequency})$\times$({\em time}) lattices. 

The $(frequency)\times (time)$ mode analog of the 2D cluster state proposed in Ref.~\cite{Larsen19} is shown in Fig.~\ref{FIG:Reprochip} (d). This can be generated by setting the splitting ratio of the BMZIs at E3, 100:0. 

Finally, we note that if the BMZIs at E2 are tuned to be 50:50, then we create a train of uncoupled ({\em frequency})$\times$({\em frequency}) quad-rail lattices described in Ref.~\cite{Wang14} and shown in Fig.~\ref{FIG:Reprochip} (e).

\subsection{Nullifiers} \label{sec:nullifiers}
An $N$-mode Gaussian pure state $\ket{\psi}$ with zero mean and complex graph $\mathbf{Z}$ can be efficiently specified by a list of $N$ linear combinations of the quadrature operators that satisfy the nullifier relation:
\begin{align}
    \hat{\mathbf{p}} - \mathbf{Z} \hat{\mathbf{q}} \ket{\psi} =0, 
\end{align}
where operators that satisfy this relation are referred to as \emph{nullifiers}~\cite{Menicucci11}, and $\hat{\mathbf{q}}=(\hat{q}_{1}, \dots, \hat{q}_{N})^{\text{T}}$, $\hat{\mathbf{p}}=(\hat{p}_{1}, \dots, \hat{p}_{N})^{\text{T}}$, and $\hat{a}_{k} = (\hat{q}_{k}+ i \hat{p}_{k})/\sqrt{2}$, where $k\in\{1,2,\cdots,N\}$.

Measuring expectation values of nullifiers plays a key role in verifying Gaussian pure states and genuine multi-partite inseparability, e.g., via the van Loock-Furusawa criterion~\cite{Loock03}. Particularly convenient are states which have nullifiers that can be re-expressed such that each only consists of either position or momentum operators. These enable particularly efficient state verification since they can be measured by setting all homodyne detectors to measure either the local position or momentum operator. 

Any state prepared from two-mode squeezed states and beamsplitters that do not mix position and momentum quadratures in the Heisenberg picture has nullifiers of this type. Explicit formula were given in Ref.~\cite{Alexander18}:
\begin{equation}
    \begin{aligned}
    &(\mathbf{I} - \mathbf{V})\hat{\mathbf{p}} \ket{\psi} \approx 0,\\
    &(\mathbf{I} + \mathbf{V})\hat{\mathbf{q}} \ket{\psi} \approx 0,
    \end{aligned}
\label{EQ:Null}
\end{equation}
where $\mathbf{I}$ is the identity operator, and $\mathbf{V}$ is the infinite squeezing limit of $\mathbf{Z}$ and is a real symmetric matrix.

\section{Architectural analysis} \label{SEC:ArchAn}
\subsection{Material considerations}
Bulk quantum-optics platforms have successfully demonstrated the generation of large-scale CV cluster states. For next-generation quantum information processing, however, issues arising from long-term stability, cost, portability, and mass productivity need be accounted for. Silicon photonics, in this regard, is a promising scalable platform as mass integration of hundreds of devices on a single chip for classical optical communication has already  been accomplished~\cite{Lipson05}. With respect to quantum information processing, silicon photonic implementations of integrated on-chip DV nonclassical sources~~\cite{Clemmen09,Sharping06,Davanco12,Takesue08}, single-photon detectors~\cite{Najafi15}, and DV logic gates~\cite{O'Brien09} have already been demonstrated. More recently, DV high-dimensional entangled states were demonstrated in a silicon-photonics platform with 500+ waveguide and interferometer components~\cite{Wang18}. Critically, quantum information processing in silicon photonics is carried out in the telecommunication band, and is thereby compatible with mature modulation, transmission, and detection technologies. Silicon, however, is not an ideal material for quantum information processing based on CVs due to its strong two-photon absorption in the telecommunication band, which precludes the generation of, e.g., highly squeezed light. Indeed, optical parametric oscillation, a key ingredient for the generation of large squeezing, has only been observed in silicon at mid-infrared~\cite{Kuyken15}, where efficient photo detectors and processing units have not yet been fully developed. 
Lithium niobate has been a widely used photonic material by virtue of its large nonlinearity and low absorption in the telecommunication window. Lithium niobate was recently employed in quantum information processing~\cite{Lenzini18,Mondain19}, but the development of quantum-information-processing platforms based on lithium niobate has been highly challenging and cost ineffective due to a lack of fabrication recipe for large-scale devices composed of hundreds to thousands of elements.

Silicon nitride (Si$_3$N$_4$), in this regard, shows its superiority in this exciting area. As a well-developed commercially-available material, Si$_3$N$_4$ has been widely used in both microelectronic and optical integrated circuits. The compatibility with the mature CMOS fabrication technology makes the Si$_3$N$_4$ platform stable, high performance, and cost effective. Unlike silicon, Si$_3$N$_4$'s ultrabroad transparency window spanning from the visible to the mid-infrared makes it immune to two-photon absorption in the telecommunication band~\cite{Moss13}. In addition, the Si$_3$N$_4$ platform enjoys three key features that render it ideal for CV quantum information processing. First, the nonlinearity of Si$_3$N$_4$ is about 20 times lower than that of silicon but the nonlinear interactions can be enhanced in ring resonators, as demonstrated in the generation of twin beams~\cite{Dutt15,Dutt16} and entangled states~\cite{Ramelow15}. Very recently, $\sim$1-dB quadrature squeezing was observed in Si$_3$N$_4$-based devices~\cite{Vaidya19,Hoff15}, opening the door to a scalable CV quantum information processing platform. Second, the Si$_3$N$_4$ platform enjoys an additional advantage in measuring frequency-multiplexed CV cluster states over the bulk quantum-optics platform based on the second-order nonlinearity: a phase-coherent soliton frequency comb produced~\cite{Kippenberg18,Bao17,Hansson14,Lamont13,Chembo13} via the third-order Kerr nonlinearity of Si$_3$N$_4$ allows for simultaneous addressing of all spectral modes of the CV cluster state. Such a capability is demonstrated in the generation of octave-spanning Kerr-soliton frequency combs in Si$_3$N$_4$~\cite{Pfeiffer17} and is unmatched by conventional bulk quantum-optics platforms in which the number of accessible spectral modes is fundamentally limited by the bandwidth of the electro-optic modulator used to produce the phase references for each spectral mode. Third, as a critical ingredient for time-multiplexed CV cluster states, long DLs of a few meters and an ultra-low loss level (0.1~dB/m) have been demonstrated in the Si$_3$N$_4$ platform~\cite{Bauters11}, representing a nearly two orders of magnitude improvement over that of silicon-based DLs. 

\subsection{Classical frequency-comb phase references}
\label{SEC:AAComb}
The generation of Kerr-soliton frequency combs has been studied extensively both in theory~\cite{Hansson14,Lamont13,Chembo13} and in experiments~\cite{Kippenberg18,Bao17,Coen13,Pfeiffer17}. Here, we provide a brief review on the generation mechanism for Kerr-soliton frequency combs, which will be subsequently used as phase references to address each spectral mode of the CV cluster state.

\subsubsection{Microring resonators}
We consider a MR with circumference $L$. In the absence of optical nonlinearities and dispersion, the resonant frequency of the cavity eigenmodes are equally spaced across the whole spectrum as shown in Fig.~\ref{FIG:Phasematch} (a). The spectral linewidth, $\kappa$, is determined by the {\em loaded} $Q$-factor, $Q^{(\text{L})}$, as $\kappa=\omega_{0}/Q^{(\text{L})}$ with $\omega_{0}=2\pi c/\lambda_{0}$, where $\lambda_{0}$ is the pump wavelength. The free spectral range (FSR) is $\Delta\omega/2\pi=c/n_{g}L$, where $n_{g}$ is the group-velocity refractive index.

\subsubsection{Kerr nonlinearity}
The Kerr effect is a third-order nonlinear phenomenon that manifests itself as a intensity-dependent refractive index $n=n_{0}+n_{2}I$, where $n_{0}$ denotes the original material refractive index, $I$ is the intensity of the field propagating in the material, and $n_2$ is the nonlinear refractive index. To study the nonlinear interactions in a MR, it is more convenient to define an $n_2$-related nonlinear coefficient
\begin{equation}
    g_0 = \frac{\hbar \omega_0^2 n_2 c}{n_0^2 V_{0}},
\end{equation}
where $\hbar = h/2\pi$ is the reduced Planck constant, $\omega_0$ is the angular frequency of the pump, $c$ is the speed of light, and $V_{0}$ is the mode volume of the MR~\cite{Chembo16}. Physically, $g_0$ quantifies the shift of the resonant frequency induced by a single pump Photonics Since $n_{2}>0$ in Si$_3$N$_4$, the resonant frequency for the pump will be red shifted relative to a cold cavity by self-phase modulation. The presence of intracavity pump power also shifts the resonant frequencies of other cavity-resonant modes via cross-phase modulation. The magnitude of cross-phase modulation is twice that of self-phase modulation, thereby leading to a doubled shift for other resonant frequencies aside from the pump, as illustrated in Fig.~\ref{FIG:Phasematch} (b). 


\subsubsection{Dispersion}
\label{SEC:Dispersion}
\begin{figure}[h]
	{\centering\includegraphics[width=7.75cm]{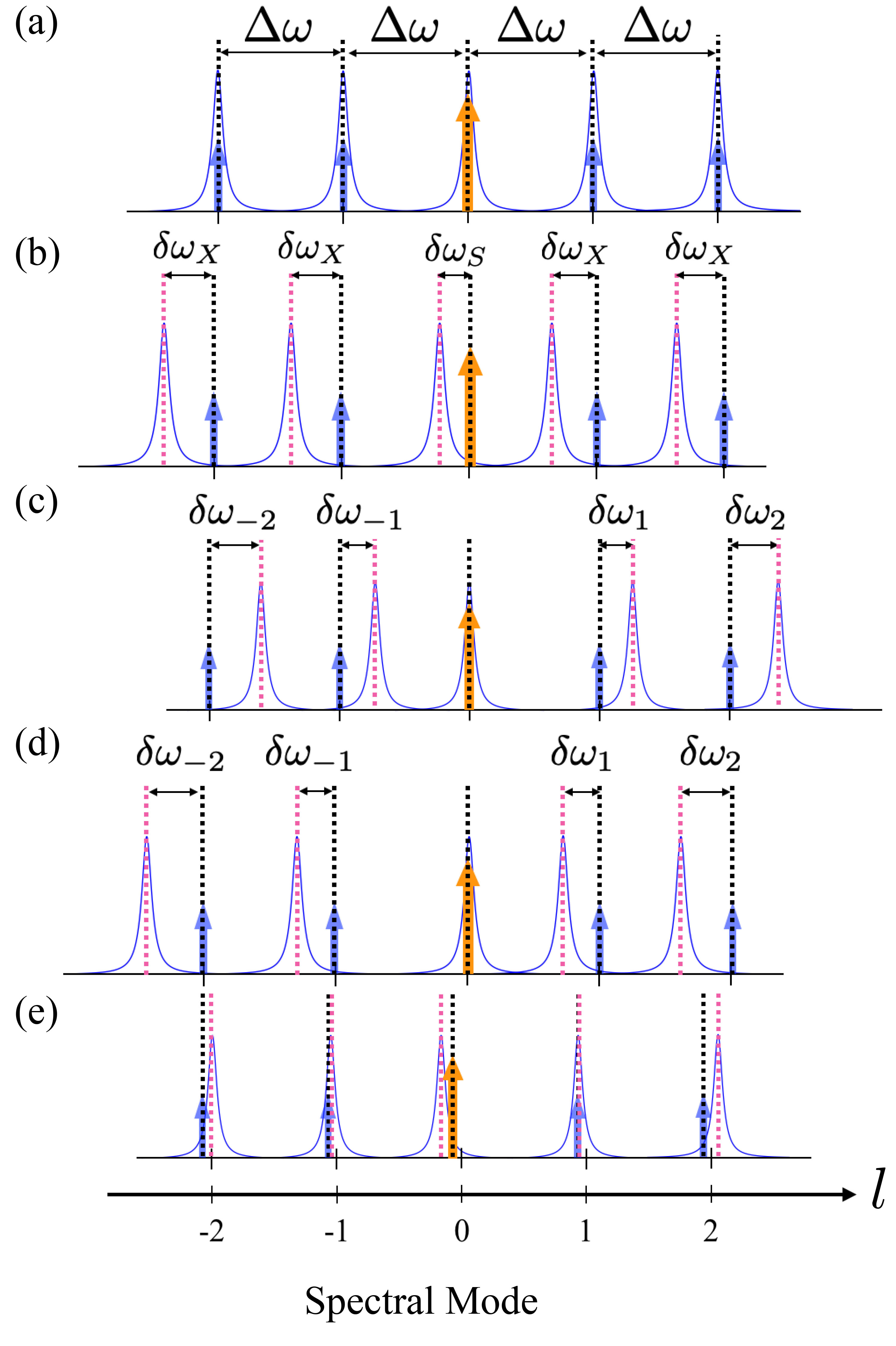}\\}
	\caption{\label{FIG:Phasematch} Shifting of cavity modes. $\delta\omega_{X}$ and $\delta\omega_{S}$ denote the frequency shifts from cross-phase modulation and self-phase modulation. $\delta \omega_{l}$ is the dispersion frequency shift of spectral mode $l$. (a) Zero-detuned ($\sigma=0$) cold cavity without considering dispersion effect ($\zeta_{2}=0$). (b) Zero-detuned hot cavity without considering dispersion effect. (c) Zero-detuned cold cavity in the case of anomalous dispersion ($\zeta_{2}>0$). (d) Zero-detuned cold cavity in the case of normal dispersion ($\zeta_{2}<0$). (e) Balancing between dispersion and Kerr nonlinearity (e.g. $\sigma<0$, $\zeta_{2}>0$). 
	}
\end{figure}

To employ MRs in broadband applications such as the generation of Kerr-soliton frequency combs or large-scale frequency-multiplexed CV cluster states, the frequency dependence of refractive index $n(\omega)$ must be accounted for. Let us first expand the frequency-dependent wavevector, $\beta(\omega) \equiv n(\omega)\omega/c$, at $\omega_{0}$ as
\begin{equation}\label{EQ:disp}
    \beta(\omega)=\sum_{s=0}^{\infty}\frac{\beta_{s}}{s!}\left(\omega-\omega_{0}\right)^s.
\end{equation}
The group velocity is $v_g = 1/\beta_1$ and $\beta_2$ is the group-velocity dispersion (GVD) parameter. Non-zero $\beta_s$'s for $s \geq 2$ lead to dispersion-induced resonant-frequency shifts. For the cavity mode indexed by $l\in\field{Z}$, the shifted resonant frequency becomes
\begin{equation}\label{EQ:cavityfreq}
    \omega_l=\omega_{0}+ l \Delta \omega + \sum_{s=2}^{\infty}\frac{\zeta_{s}}{s!}l^s,
\end{equation}
where $\zeta_{s}$ can be derived from $\beta_s$'s and $\Delta \omega$, as shown in Appendix~\ref{SEC:Disp}. The shifted resonant frequencies are illustrated in Fig.~\ref{FIG:Phasematch} (c) for $\zeta_{2}>0$, i.e., anomalous dispersion and Fig.~\ref{FIG:Phasematch} (d) for $\zeta_{2}<0$, i.e., normal dispersion. A strong pump red-shifts the resonant frequencies, while the anomalous dispersion blue-shifts the resonant frequencies. The overall frequency shift is balanced to ensure that a large number of spectral modes reside approximately on the cavity resonances, as shown in Fig.~\ref{FIG:Phasematch} (e). This is a key to achieving efficient Kerr-soliton and CV cluster-state generation~\cite{Hansson14,Lamont13,Chembo13,Kippenberg18,Bao17,Coen13,Pfeiffer17}. 

\subsubsection{Classical dynamics}
\label{SEC:CD}
We now formulate the generation of Kerr-soliton frequency combs that will serve as phase references for the CV cluster states. We consider an MR pumped by a single mode situated at $l = 0$ with amplitude $A_{\text{in}}$ (in units of $\sqrt{\rm photons/s}$). Let the intracavity field of spectral mode $l$ be $A_l$ (in units of $\sqrt{\rm photon\,number}$). The evolution of the intracavity modes is governed by the coupled-mode equations~\cite{Chembo 10,Chembo13,Chembo10,Chembo16,Herr13}:
\begin{equation}\label{EQ:CMfield}
\begin{aligned}
    \frac{d{A}_{l}(t)}{dt}&=-\frac{\kappa}{2} A_{l}(t)+i\left(\sigma-\sum_{s=2}^{\infty}\frac{\zeta_{s}}{s!}l^s\right)A_{l}(t)\\
    &+ig_{0}\sum_{j,k}A_{j}(t) A_{k}^{*}(t) A_{k+l-j}(t)+\delta_{l,0}\sqrt{\kappa^{(\text{o})}}\,A_\text{in}(t).
\end{aligned}
\end{equation}
Here, $\kappa=\kappa^{(\text{i})}+\kappa^{(\text{o})}$ is the spectral linewidth, which describes the total power decay rate. It accounts for the intrinsic cavity loss $\kappa^{(\text{i})}$ and the out-coupling loss $\kappa^{(\text{o})}$; $\sigma$ is the pump detuning away from pump frequency $\omega_0$. On the right-hand side of Eq.~(\ref{EQ:CMfield}), the first term corresponds to intracavity power decay, the second term relates to the resonant frequency shift due to dispersion, the third term describes the nonlinear Kerr interactions between different cavity modes, including self-phase modulation, cross-phase modulation, and FWM, and the last term links the extracavity pump with the intracavity field. The coupled-mode equations represent a {\em frequency-domain} approach in which the evolution of each spectral mode is derived. Alternatively, the classical dynamics can be studied in the {\em time domain} by the Lugiato-Lefever equation (LLE)~\cite{Coen13,Bao17,Hu17}:
\begin{equation}
\begin{aligned}
    t_{R}\frac{\partial E(t)}{\partial t}=&\left[-\frac{\alpha+\theta}{2}-i\delta+iL\sum_{s=2}^{\infty}\frac{\beta_{s}}{s!}\left(i\frac{\partial}{\partial \tau}\right)^{s}\right]E(t)\\
    &+i\gamma L|E(t)|^2E(t)+\sqrt{\theta}\;E_{\text{in}}(t).
\end{aligned}
\end{equation}
Here, $t_{R}=2\pi/\Delta\omega$ is the round trip time, $E(t)$ describes the intracavity field involving all cavity modes (in units of $\sqrt{\text{W}}$). $t$ and $\tau$ denote the {\em slow} time and {\em fast} time of the fields, $\alpha=\omega_{0}t_{R}/Q^{(\text{i})}$ is the normalized intrinsic cavity loss, where $Q^{(\text{i})}$ is the {\em intrinsic} $Q$-factor. $\theta$ is the normalized out-coupling loss. The normalized total cavity loss encompassing both the intrinsic and out-coupling contributions is $\alpha+\theta=\omega_{0}t_{R}/Q^{(\text{L})}$. $\delta$ is the normalized pump detuning, $\gamma= n_{2}\omega_{0}L/cV_{0}$ is the effective nonlinear coefficient, and $E_{\text{in}}$ is the input pump field.

We simulated the formation of Kerr-soliton frequency combs in the overcoupling regime using the LLE. In our simulation, we consider a MR circumference $L=15.7$ mm, $Q^{(\text{L})} =2\times10^{6}$, and $Q^{(\text{i})}=2.22\times10^{7}$~\cite{Ji17,Xuan16}. The pump wavelength is chosen to be $\lambda_{0}=1549.6$ nm. To design the MR waveguide with the desired dispersion properties, we utilized the simulation environment COMSOL supplied with Sellmeier equations for Si$_3$N$_4$ reported in Ref.~\cite{Zhang14}. The designed MR waveguide has a rectangular cross-section with width $W_{\rm C}=2.1$ $\muup$m and height $H_{\rm C}=0.82$ $\muup$m. The simulation result gives the effective refractive indices $n_{0}=1.85$ and $n_{g}=2.05$ and dispersion coefficients, as shown in Table~\ref{TAB:HOD}. 
\begin{center}
\begin{table}[h]
\begin{tabular}{ |c|c| }
\hline
$\beta_{2}$ (s$^2$/m) & $-1.986\times10^{-25}$ \\ 
\hline
$\beta_{3}$ (s$^3$/m) & $2.546\times10^{-39}$ \\ 
\hline
$\beta_{4}$ (s$^4$/m) & $3.318\times10^{-52}$ \\ 
\hline
$\beta_{5}$ (s$^5$/m) & $-1.625\times10^{-65}$ \\ 
\hline
$\beta_{6}$ (s$^6$/m) & $-3.863\times10^{-79}$ \\ 
\hline
$\beta_{7}$ (s$^7$/m) & $-4.000\times10^{-92}$ \\ 
\hline
$\beta_{8}$ (s$^8$/m) & $-7.916\times10^{-106}$ \\ 
\hline
\end{tabular}
\caption{High order dispersions with waveguide geometry, $W_{\rm C}=2.1$ $\muup$m and $H_{\rm C}=0.82$ $\muup$m.}
\label{TAB:HOD}
\end{table}
\end{center}
The nonlinear index is $n_{2}=2.5\times10^{-19}$ m$^2$/W~\cite{Levy 10}, corresponding to an effective nonlinear coefficient of $\gamma=0.59$ (1/Wm). Also, the simulated effective refractive index determines an FSR of $\Delta \omega/2\pi = 9.32$ GHz.

\begin{figure}[h]
	{\centering\includegraphics[width=1\linewidth]{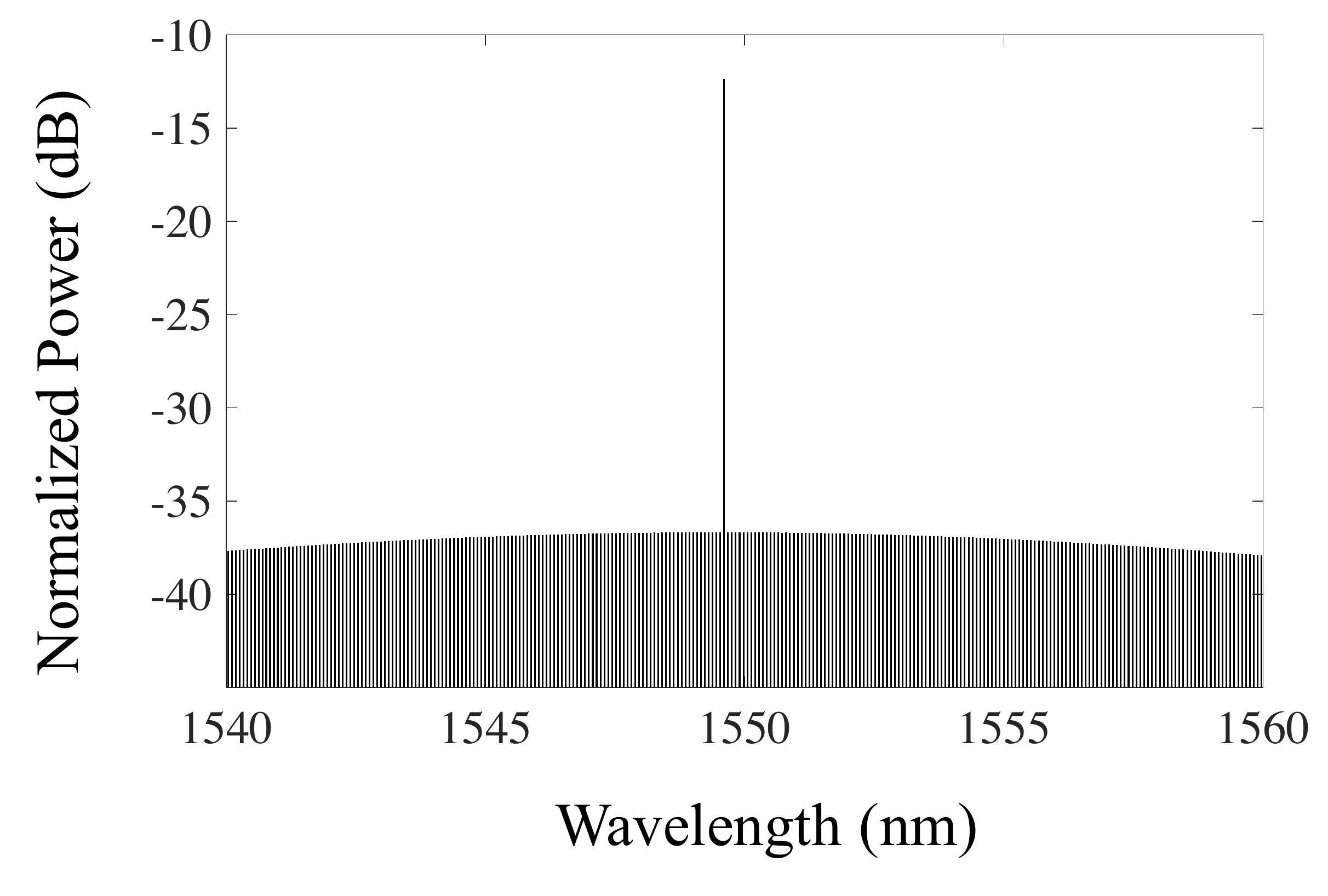}\\}
	\caption{\label{FIG:SolitonSpec} The spectrum of a stable Kerr-soliton frequency comb. Each frequency tooth can serve as a pump or a phase reference that addresses a corresponding spectral mode in the CV cluster state.
	}
\end{figure}

To produce Kerr-soliton frequency combs, the MR is pumped above its oscillating threshold $P_{\text{th}}=51.53$ mW by a c.w. pump with a power level of $P_{\text{in}} =1.2$ W and an initial normalized pump detuning of $\delta = 0$. Subsequently, the pump detuning is adjusted to 0.21, 0.42 and 0.75 at, respectively, 25 ns, 50 ns and 75 ns, when a stable Kerr-soliton is observed, as plotted in Fig.~\ref{FIG:SolitonSpec} its spectrum.

\subsection{Quantum dynamics}
To study the quantum dynamics, in particular, the formation of entanglement between different spectral modes, the classical coupled-mode equations need to be augmented with quantum field operators. Specifically, the quantum field operator $\hat{a}_{l}(t)$ for the $l$-th intracavity field can be decomposed into a classical mean field $A_l(t)$ and a quantum fluctuation operator $\delta\hat{a}_{l}(t)$~\cite{Chembo13}:
\begin{equation}
    \hat{a}_{l}(t)=A_{l}(t)+\delta \hat{a}_{l}(t). 
    \label{eq:modequcl}
\end{equation}

The evolution of $A_l(t)$ is derived using the classical coupled-mode equations in Eq.~\ref{EQ:CMfield}, while the dynamics of $\delta\hat{a}_{l}(t)$ is governed by quantum coupled-mode equations:
\begin{equation}
\label{EQ:quantum_coupled_mode}
\begin{aligned}
    \frac{d\delta \hat{a}_{l}(t)}{dt} =& -\left[\frac{\kappa}{2}-i\left(\sigma-\sum_{s=2}^{\infty}\frac{\zeta_{s}}{s!}l^{s}\right)\right]\delta \hat{a}_{l}(t)\\
    &+ig_{0}|A_{0}|^{2}\delta\hat{a}_{-l}^{\dagger}(t)+\sum_{\text{s=i,o}}\sqrt{\kappa^{(\text{s})}}\;\hat{\mathcal{V}}^{(\text{s})}_{l}(t),
\end{aligned}
\end{equation}
where $\kappa=\omega_{0}/Q^{(\text{L})}$, $A_{0}$ is the stabilized classical field at the pump mode. Here, without loss of generality, we assume the pump mode is situated at $l=0$. $\hat{\mathcal{V}}^{(\text{i})}_{l}(t)$ and $\hat{\mathcal{V}}^{(\text{o})}_{l}(t)$ are the vacuum noise field operators at the spectral mode $l$, induced by the cavity intrinsic loss and the out-coupling loss, respectively. The introduction of the vacuum noise operators is required to preserve the Heisenberg uncertainty principle~\cite{Haus95}. Both noise operators satisfy the commutation relations:
\begin{equation}
    \label{EQ:CR}
\begin{aligned}
    \lbrack\hat{\mathcal{V}}^{(\text{i})}_{l}(t),\hat{\mathcal{V}}^{\dagger (\text{i})}_{l'}(t')\rbrack&=\delta_{l,\,l'}\delta\left(t-t'\right),\\
    \lbrack\hat{\mathcal{V}}^{(\text{o})}_{l}(t),\hat{\mathcal{V}}^{\dagger (\text{o})}_{l'}(t')\rbrack&=\delta_{l,\,l'}\delta\left(t-t'\right).
\end{aligned}
\end{equation}

To study the quantum dynamics, it is convenient to derive the spectrum of the quantum field operators. To do so, we take the Fourier transform on both sides of Eq.~(\ref{EQ:quantum_coupled_mode}) and obtain
\begin{equation}
\begin{aligned}
    \label{EQ:FCoupled}
    -i\omega\; \delta\hat{\tilde{a}}_{l}(\omega) =& -\left[\frac{\kappa}{2}-i\left(\sigma-\sum_{s=2}^{\infty}\frac{\zeta_{s}}{s!}l^{s}\right)\right]\delta\hat{\tilde{a}}_{l}(\omega)\\
    &+ig_{0}A_{0}^2\;\delta\hat{\tilde{a}}^{\dagger}_{-l}(\omega)+\sum_{\text{s=i,o}}\sqrt{\kappa^{(\text{s})}}\;\hat{\tilde{\mathcal{V}}}^{(\text{s})}_{l}(\omega),
\end{aligned}
\end{equation}
where ``$\sim$'' on the top denotes the frequency-domain operators obtained by taking Fourier transform on the time-domain operators. 

The intracavity field is coupled out to the bus waveguide to form the out-coupling field residing in the bus waveguide, which can be directly measured and characterized. Here, the out-coupling field is represented as $\delta\hat{\tilde{a}}^{(\text{out})}$, which relates to the intracavity field, $\delta\hat{\tilde{a}}_{l}$, via
\begin{equation}
\label{EQ:Output}
    \delta\hat{\tilde{a}}_{l}^{(\text{out})}(\omega)=\sqrt{\kappa^{(\text{o})}}\;\delta\hat{\tilde{a}}_{l}(\omega)-\hat{\tilde{\mathcal{V}}}_{l}^{(\text{o})}(\omega).
\end{equation}

From Eq.~(\ref{EQ:FCoupled}) and Eq.~(\ref{EQ:Output}), $\delta\hat{\tilde{a}}^{(\text{out})}_{l}(\omega)$ and $\delta\hat{\tilde{a}}^{\dagger (\text{out})}_{-l}(\omega)$, are written in a matrix-form representation,
\begin{equation}\label{EQ:extfield_freq}
    \begin{aligned}
    \begin{pmatrix}
    \delta \hat{\tilde{a}}^{(\text{out})}_{l}(\omega)\\ \delta \hat{\tilde{a}}^{\dagger (\text{out})}_{-l}(\omega)
    \end{pmatrix}=&-M^{-1}_{l}
        \sum_{\text{s=i,\,o}}\sqrt{\kappa^{(\text{s})}\kappa^{(\text{o})}}
        \begin{pmatrix}
        \hat{\tilde{\mathcal{V}}}^{(\text{s})}_{l}(\omega)\\
        \hat{\tilde{\mathcal{V}}}^{\dagger (\text{s})}_{-l}(\omega)
        \end{pmatrix}\\
        &-\begin{pmatrix}
        \hat{\tilde{\mathcal{V}}}^{(\text{o})}_{l}(\omega)\\
        \hat{\tilde{\mathcal{V}}}^{\dagger (\text{o})}_{-l}(\omega)
        \end{pmatrix}.
    \end{aligned}
\end{equation}
Here, 
\begin{equation}
\begin{aligned}
    M_{l} &=\begin{pmatrix}
    J_{l}+i\omega && ig_{0}A^2_{0} \\
    -ig_{0}A_{0}^{* 2} && J_{-l}^{*}+i\omega
    \end{pmatrix},\\
    J_{\pm l}&=i\left[\sigma-\sum_{s=2}^{\infty}\frac{\zeta_{s}}{s!}\left(\pm l\right)^s+2g_{0}|A_0|^2\right]-\frac{\kappa}{2}.
    \end{aligned}
\end{equation}
The introduction of IBSs, DLs and waveguide crossings causes attenuation on the power of the extracavity quantum fields by a factor of $1-\eta$. To account for the power attenuation, we introdue an attenuated quantum-field operator, $\delta\hat{\tilde{a}}^{(\text{att})}_{l}$, modeled by
\begin{equation}
    \label{EQ:OUT}
    \delta\hat{\tilde{a}}^{(\text{att})}_{l}(\omega)=\sqrt{\eta}\;\delta\hat{\tilde{a}}^{(\text{out})}_{l}(\omega)+\sqrt{1-\eta}\;\hat{\tilde{\mathcal{V}}}^{(\text{a})}_{l}(\omega),
\end{equation}
where $\eta\in\left[0,1\right]$, and $\hat{\tilde{\mathcal{V}}}^{(\text{a})}_{l}(\omega)$ is the vacuum noise operator associated with the power attenuation. $\hat{\tilde{\mathcal{V}}}^{(\text{a})}_{l}(\omega)$ and its corresponding noise operator in the time domain, $\hat{\mathcal{V}}^{(\text{a})}_{l}(t)$, satisfy the commutation relations similar to that of Eq.~(\ref{EQ:CR}). 

From Eq.~(\ref{EQ:OUT}), the nullifiers~\cite{Pysher11,Chen14} for verifying the multipartite inseparability of the 0D, 1D, 2D, and 3D CV cluster states can be derived~\cite{Loock03}. These follow immediately from the graphical representation of the state and Eqs.~(\ref{EQ:Null}).


To derive the nullifiers, recall the definition of position and momentum operators ($\hat{q}$ and $\hat{p}$):
\begin{equation}
\begin{aligned}
    \hat{q}&=\frac{1}{\sqrt{2}}\left(\hat{a}+\hat{a}^{\dagger}\right),\\
   \hat{p}&=\frac{1}{i\sqrt{2}}\left(\hat{a}-\hat{a}^{\dagger}\right),
\end{aligned}
\end{equation}
where $\hat{a}$ and $\hat{a}^{\dag}$ are the annihilation and creation operators, respectively, applied to any dimension from zero to three.
Similarly, we can define the rotated quadrature operators as 
\begin{equation}
    \begin{aligned}
        \hat{q}(\theta)&= \hat{q}\cos{\theta}-\hat{p}\sin{\theta},\\
        \hat{p}(\theta)&=\hat{q}\sin{\theta}+\hat{p}\cos{\theta}.
    \end{aligned}
\end{equation}

This $\theta$ parameter arises due to the phase difference between the local oscillator and the quantum fields. By tuning $\theta\in\left[0,2\pi\right)$, we search for the angle that results in the maximum squeezing of the nullifier variance.

\begin{figure*}
	{\centering\includegraphics[width=0.75\linewidth]{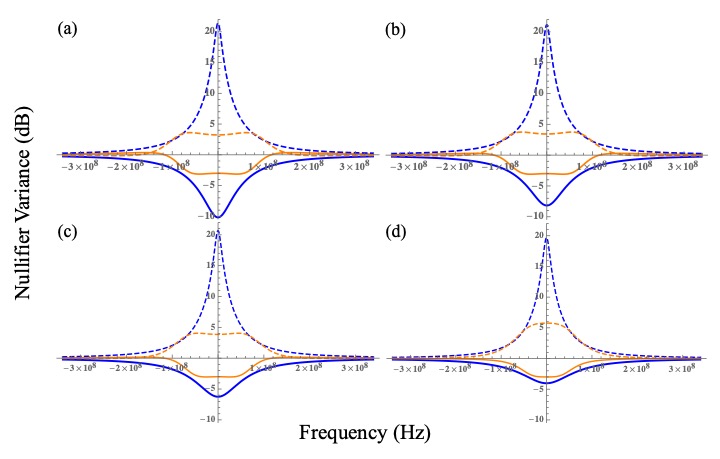}\\}
	\caption{\label{FIG:Spectra} The squeezing and anti-squeezing spectra, normalized to the shot-noise limit, for (a) 0D, (b) 1D, (c) 2D, and (d) 3D CV cluster states . Blue solid (dashed) curves are the squeezing (anti-squeezing) spectra with the maximum squeezing levels in 10.18~dB, 8.17~dB, 6.25~dB, and 4.03~dB in 0D, 1D, 2D and 3D cases. Orange solid (dashed) curves are the squeezing (anti-squeezing) spectra at 3~dB squeezing.}
\end{figure*}

Fig.~\ref{FIG:Spectra} shows the simulation result for nullifier variances, under the physical parameters $\beta_{2}$, $L$, $\lambda_{0}$, $Q^{(\text{L})}$, $Q^{(\text{i})}$, $n_{2}$, $n_{0}$ and $V_{0}$ specified in Sec.~\ref{SEC:Design}.

\subsubsection{Generation of 0D CV cluster states}
In the 0D case, the output quantum fields form pair-wise two-mode squeezed states between mode $l$ and $-l$ with the nullifiers
\begin{equation}\label{EQ:twomodenull}
    \begin{aligned}
        &\delta\hat{\tilde{q}}_{l}(\omega)-\delta\hat{\tilde{q}}_{-l}(\omega),\\
        &\delta\hat{\tilde{p}}_{l}(\omega)+\delta\hat{\tilde{p}}_{-l}(\omega).
    \end{aligned}
\end{equation}
The mode configuration is illustrated in Fig.~\ref{FIG:0DCluster}. We consider $l\leq l_{\rm 3dB}$, where $\pm l_{\rm 3dB}$ index the entangled spectral modes at which the squeezing level is right above 3~dB. The optimal squeezing spectrum is displayed in Fig.~\ref{FIG:Spectra} (a) and shows a 10.18~dB squeezing level.

\subsubsection{Generation of 1D CV cluster states}
To generate 1D cluster states, we prepare two 0D cluster states produced at spatial modes `a' and `b'. The two 0D cluster states are subsequently mixed through a 50:50 IBS shown explicitly in Fig.~\ref{FIG:1DCluster}. In passing through the 50:50 IBS, quantum fields are linearly processed, leading to the nullifiers,
\begin{equation}\label{EQ:1DNull}
    \begin{aligned}
        &\frac{\delta\hat{\tilde{q}}_{\text{a},\,l}(\omega)-\delta\hat{\tilde{q}}_{\text{a},-l}(\omega)}{\sqrt{2}}+\frac{\delta\hat{\tilde{q}}_{\text{b},\,l}(\omega)-\delta\hat{\tilde{q}}_{\text{b},-l}(\omega)}{\sqrt{2}},\\
        &\frac{\delta\hat{\tilde{q}}_{\text{a},\,l}(\omega)-\delta\hat{\tilde{q}}_{\text{a},4-l}(\omega)}{\sqrt{2}}-\frac{\delta\hat{\tilde{q}}_{\text{b},\,l}(\omega)-\delta\hat{\tilde{q}}_{\text{b},4-l}(\omega)}{\sqrt{2}},\\
        &\frac{\delta\hat{\tilde{p}}_{\text{a},\,l}(\omega)+\delta\hat{\tilde{p}}_{\text{a},-l}(\omega)}{\sqrt{2}}+\frac{\delta\hat{\tilde{p}}_{\text{b},\,l}(\omega)+\delta\hat{\tilde{p}}_{\text{b},-l}(\omega)}{\sqrt{2}},\\
        &\frac{\delta\hat{\tilde{p}}_{\text{a},\,l}(\omega)+\delta\hat{\tilde{p}}_{\text{a},4-l}(\omega)}{\sqrt{2}}-\frac{\delta\hat{\tilde{p}}_{\text{b},\,l}(\omega)+\delta\hat{\tilde{p}}_{\text{b},4-l}(\omega)}{\sqrt{2}},
    \end{aligned}
\end{equation}
where the sub-indices denote "\emph{spatial mode}" and "\emph{spectral mode}". IBS can operate across a wide frequency range from 1500 nm to 1580 nm, while introducing an estimated insertion loss of 0.28~dB/IBS~\cite{Zhang13}. The output squeezing spectrum is displayed in Fig.~\ref{FIG:Spectra} (b) with a squeezing level of 8.17~dB.

\subsubsection{Generation of 2D CV cluster states}
To generate 2D CV cluster states, 1D cluster states are processed by linear optics~\cite{Menicucci07}, as shown in Fig.~\ref{FIG:2DCluster}. With the introduction of the DL, the quantum-field operators acquire an additional index, time. We denote the set of temporal labels as $T=\{t_{j} \: | \; t_{j}=t_{0}+j\; \delta t, \; j\in\field{Z}^{+}_{0}\}$, where $t_{0}$ is the starting time. The number of temporal modes is determined by the amount of time over which the experiment is executed. The DL spreads the momentary entanglement to a series of entangled temporal modes with spacing $\delta t$. The nullifiers can be derived as the linear combinations of the fields, $\delta\hat{\tilde{a}}_{s,\,l,\,t_{n}}$, where $s\in\{\text{a}, \text{b}\}$, $l\leq l_{3\rm dB}$, and $t_{n}\in T$, are fully described by Eq.~\ref{EQ:Null}. The corresponding graph state is further displayed in the stage $(iv)$ of Fig.~\ref{FIG:2DCluster}. 

The generation of 2D cluster states requires two 50:50 IBSs, one UMZI, and one DL. Overall, the insertion loss of the IBSs and the propagation loss of DL need be considered. Reported propagation loss of DLs was as low as 0.1~dB/m~\cite{Bauters11}. Taking into account all losses, the output squeezing level becomes 6.25~dB, as shown in Fig.\ref{FIG:Spectra} (d).

\subsubsection{Generation of 3D CV cluster states}
To generate 3D CV cluster states, we replicate two sets of the 2D cluster state setup and process the output fields based on the structure illustrated in Fig.~\ref{FIG:Programmable}. At each time step, the state consists of a 2D entangled structure, with axes $f_{1}$ and $f_{2}$ made by appropriate frequency multiplexing. To achieve this, we choose our four input pump spectral modes to be $l=0,2,1-\Delta,1+\Delta$, where the total number of spectral modes is determined by the phase-matching bandwidth, and $\Delta$ specifies how the total number of frequency indices are split between the two lattice axes $f_{1}$ and $f_{2}$~\cite{Wang14}. The nullifiers of 3D cluster state are shown in stage $(v)$ of Fig.~\ref{FIG:Programmable} and described by Eq.~\ref{EQ:Null}.

Making the structure depicted in Fig.~\ref{FIG:Programmable} in a small chip requires some of the waveguides to cross others. To examine the squeezing level in the 3D cluster case, we need to account for the loss arising from waveguide crossing, aside from insertion loss and propagation loss. The experimentally achievable crossing loss can be as low as 0.015~dB/cross~\cite{Blumenthal18}. Thus, the overall attenuation results in an output squeezing level of 4.03~dB, as shown in Fig.~\ref{FIG:Spectra} (d).

\subsection{Effect of dispersion}
\label{SEC:EffectDispersion}
To take into account the effects of dispersion, we model the cross-section of each quantum MR as a rectangle, with width $W_{\rm Q}$ and height $H_{\rm Q}$, similar to that of the classical MR in Sec.~\ref{SEC:CD}. Tuning the aspect ratio $W_{\rm Q}/H_{\rm Q}$ allows us to obtain different dispersion parameters, which in turn determine the dimensions of the cluster state.

\begin{figure}[h]
	{\centering\includegraphics[width=1\linewidth]{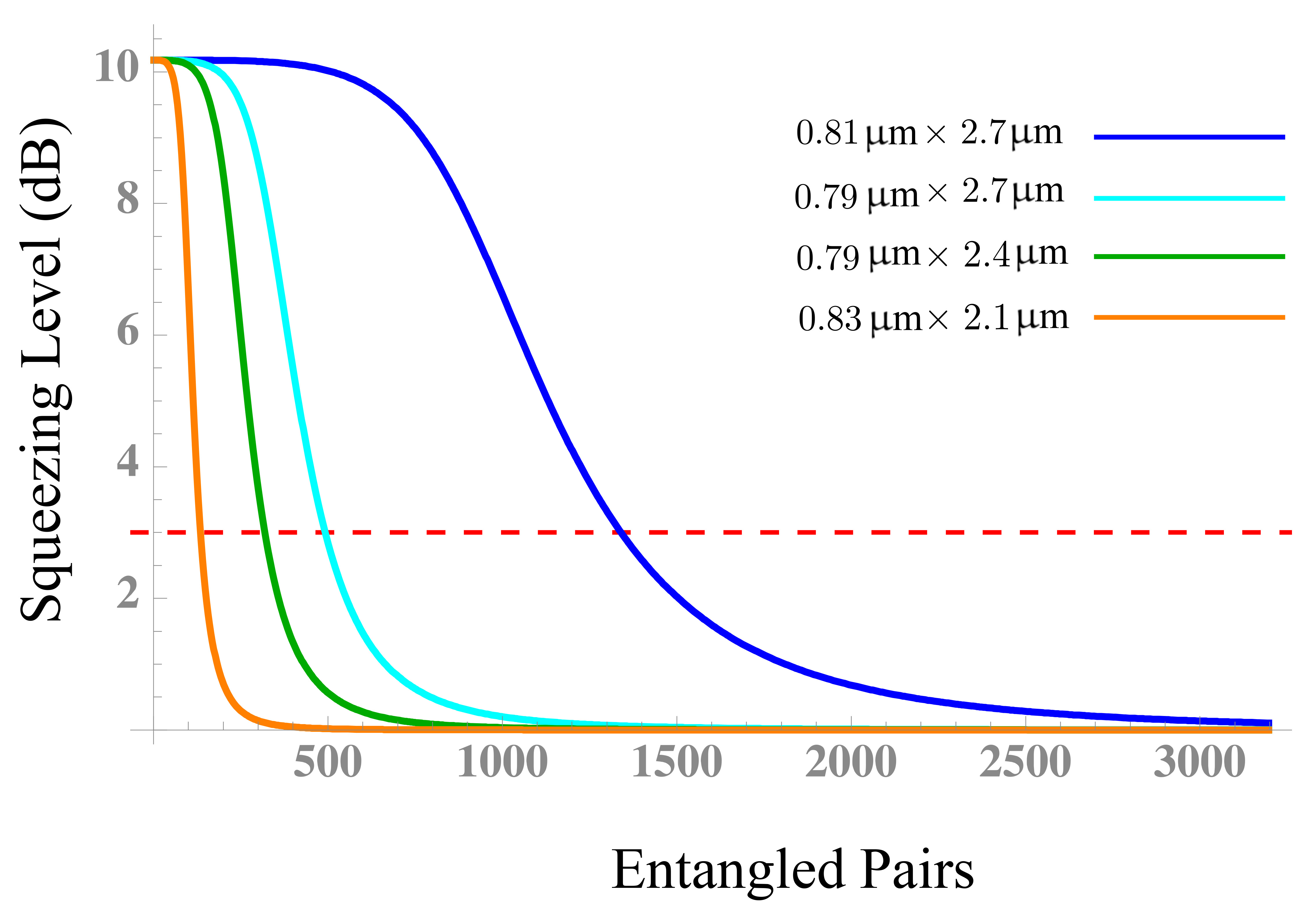}\\}
	\caption{\label{FIG:Entangledmode} Normalized squeezing levels for the 0D case versus the entangled pair index for four waveguide cross section configurations. Red dashed horizontal line is at the 3~dB squeezing level. The numbers of entangled pairs above the 3~dB squeezing level for the four cases are $l_{3\rm dB }=1340,500,331,143$.}
\end{figure}

Given $W_{\rm Q}$, $H_{\rm Q}$, and the pump wavelength $\lambda_{0}=1549.6$ nm, dispersion parameters $\beta_{1}$, $\beta_{2}$, and $\beta_{3}$ can be calculated by COMSOL. Four combinations of $W_{\rm Q}$ and $H_{\rm Q}$ are listed in Table~\ref{TAB:Disp}. The 0D-cluster-state squeezing levels as a function of available mode pairs for the four settings are shown in Fig.~\ref{FIG:Entangledmode}.

The wavevector mismatch at spectral modes $l$ and $-l$, $\Delta\beta_{l;-l}=\beta_{l}+\beta_{-l}-2\beta_{0}$, is written as
\begin{equation}
\label{EQ:Phasemismatch}
    \Delta\beta_{l;-l}\approx\frac{1}{c}\left(\zeta_{2}l^2-2g_{0}|A_{0}|^2\right).
\end{equation}
Since $\Delta\beta_{l;-l}$ ties to the amount of squeezing, Eq.~(\ref{EQ:Phasemismatch}) indicates that the size of the cluster state is determined by $\beta_{2}$. A lower $\beta_{2}$ results in greater number of entangled spectral modes, i.e., a larger $l_{\rm 3dB}$, as depicted in Fig.~\ref{FIG:Entangledmode}.

\begin{center}
\begin{table}[h]
\begin{tabular}{ |c|c||c|c|c| } 
\hline
 $H_{\rm Q}$ & $W_{\rm Q}$ & $\beta_{1}$ ($\times10^{-9}$) & $\beta_{2}$ ($\times10^{-26}$)  & $\beta_{3}$ ($\times10^{-39}$)\\
\hline
 $\muup$m  & $\muup$m  & s/m & s$^2$/m & s$^3$/m\\ 
\hline
 0.81 & 2.7 & 6.834 & $-$0.133 & 0.335 \\ 
\hline
 0.79 & 2.7 & 6.837& $-$0.988 &$-$1.090 \\ 
\hline
 0.79 & 2.4 & 6.850 & $-$2.359 & $-$0.383 \\
\hline
 0.83 & 2.1 & 6.865 & $-$13.27 & 1.226 \\ 
\hline
\end{tabular}
\caption{Selected four combinations of $W_{\rm Q}$ and $H_{\rm Q}$ and their corresponding $\beta_{1}$'s, $\beta_{2}$'s, and $\beta_{3}$'s.}
\label{TAB:Disp}
\end{table}
\end{center}

\subsection{Experimental realization}
\label{SEC:Design}
In this section we summarize the parameters for the CV cluster-state platform, estimate the size and squeezing of the resultant state, and provide further details about the experimental realization. 

Given the dispersion parameters in Table~\ref{TAB:Disp}, a quantum MR with $W_{\rm Q}=2.7$ $\muup$m and $H_{\rm Q}=0.81$ $\muup$m is chosen so that the dispersion parameters, $\beta_{1}$, $\beta_{2}$, and $\beta_{3}$, result in a large number of entangled spectral modes. In the following, we focus on improving the squeezing level, by optimizing other physical parameters of the quantum MR.
 
The pump detuning is selected to be $\sigma/2\pi=-82$ MHz apart from the pump at $\sim$193 THz. The input pump power is $P_{\text{in}}=55.63$ mW, below the threshold $P_{\text{th}}=65.45$ mW. The length of the DLs is designed to be $L_{\text{DL}}\geq v_{g}\Delta T=151.27$ cm, where $\Delta T=2\pi/\kappa=10.34$ ns. In our designed quantum MR, we set its {\em loaded} $Q$-factor $Q^{(\text{L})}=2\times10^6$, {\em intrinsic} $Q$-factor $Q^{(\text{i})}=2.22\times10^{7}$, and FSR $\Delta\omega/2\pi=9.32$ GHz, so that these parameters match those of the classical MR described in Sec.~\ref{SEC:CD}. By doing so, we can utilize the frequency comb, from the classical MR, to address the CV cluster state, from quantum MR, in homodyne detection.

The above physical parameters lead to maximum squeezing levels of 10.18~dB, 8.17~dB, 6.25~dB and 4.03~dB for 0D, 1D, 2D, and 3D cluster states, respectively. We estimate the size of CV cluster states for each dimension assuming a 3~dB cutoff for modes in the frequency direction. For the states displayed in Figs.~\ref{FIG:0DCluster}, \ref{FIG:1DCluster}, \ref{FIG:2DCluster}, and \ref{FIG:Programmable}, these are
\begin{equation}
\label{EQ:SIZE}
\begin{aligned}
    N^{(0 \text{D})}&=1\times2680,\\
    N^{(1 \text{D})}&=2\times2604,\\
    N^{(2 \text{D})}&=2\times2472\times\tau,\\
    N^{(3 \text{D})}&=4\times45\times45\times\tau,
\end{aligned}
\end{equation}
respectively, which follow the multiplication forms as
\begin{equation}
\begin{aligned}
    N^{(0 \text{D})}\rightarrow&\,(\text{spatial \#})\times(\text{spectral \#}), \\
    N^{(1 \text{D})}\rightarrow&\,(\text{spatial \#})\times(\text{spectral \#}),\\
    N^{(2 \text{D})}\rightarrow&\,(\text{spatial \#})\times(\text{spectral \#})\times(\text{temporal \#}),\\
    N^{(3 \text{D})}\rightarrow&\,(\text{spatial \#})\times(\text{spectral \#})_{f_{1}}\times(\text{spectral \#})_{f_{2}}\times\\
    &(\text{temporal \#}).
\end{aligned}
\end{equation}
In the 3D case, we choose $\Delta\approx O(\sqrt{N^{(3\text{D})}/4 \tau})$ so that the number of spectral modes in $f_{1}$ agrees with that of $f_{2}$. Overall, our scheme does not set an upper bound for the temporal mode index $\tau$, and, therefore, $N^{(2 \text{D})}$ and $N^{(3 \text{D})}$ can, in principle, be extended to infinity.

Thus, our approach should provide an experimentally feasible scheme to generate time-frequency multiplexed cluster states in a photonic circuit. Nevertheless, in a real experiment, there are still some challenges to be overcome. In the following, we list one primary challenge along with its solution.

To generate large-scale CV cluster states, we need to prepare several identical MRs---2 MRs for the 1D and 2D cases, and 4 MRs for the 3D case. However, fabricating several effectively identical MRs poses an engineering challenge. Fabrication errors may, for example, result in variations in the FSR of each MR and ultimately a reduction in the quality of the output CV cluster states. One way to overcome this problem is to sandwich each MR by two parallel bus waveguides. Then, we send a pair of pump fields from each bus waveguides in counter-propagating directions so that they are into the same MR base. This allows for the FSRs of the MRs to be matched by the thermo-optical fine tuning.

\section{Quantum computing with the CV cluster state}
\label{SEC:QC}
In previous sections, we provided details for how to generate 0D, 1D, 2D, and 3D CV cluster states. Here we describe how such states can be used for one-way quantum computing.

Implementing one-way quantum computing requires homodyne measurements of each spectral-temporal mode of the CV cluster state. The local oscillators required for homodyne detection can be generated via classical frequency combs from a supplementary MR system. We pump the supplementary MR above threshold to  experimentally realize optical soliton generation by choosing the physical parameters in Sec.~\ref{SEC:CD}. The frequency teeth of the generated optical are coherent, nearly equidistant from each other, and can be a new source of classical fields or serve as multiple phase references. To implement independently tunable homodyne detection on multiple spectral modes, the relative phase of each tooth must be variable. This can be implemented using a waveshaper~\cite{Xu18}.

Though we are primarily interested in describing quantum computing with the 3D cluster state, we first briefly summarize what is known about the other cases. The 0D case is not sufficiently connected for use in one-way quantum computing. The 1D case is a resource for single-mode one-way quantum computing, as described in Ref.~\cite{Alexander14}. The 2D case is a universal resource, and can implement multimode gates via the one-way quantum-computing protocol described in Ref.~\cite{Alexander16}. The quantum circuits that can be implemented on this resource are local in (1+1) dimensions.   Below, we provide a one-way quantum-computing protocol for the 3D resource state. Our protocol is capable of implementing local quantum circuits in (2+1) dimensions, which should improve quantum circuit compilation relative to 2D resources. 

Recall also that the states described in the previous sections were technically not CV cluster states. This does not cause any issues because the generation circuit only consists of two-mode squeezing and 50:50 beamsplitters that do not mix the position part and momentum part of each quadrature. For this reason, the states generated are equivalent to CV cluster states by application of a {$\pi/4$-phase} delay on all modes. This change can be incorporated into the measurement device. See Appendix.~\ref{sec:graphnot} and Ref.~\cite{Alexander18} for more details. For the procedure described below, we assume that these phase delays have already been implemented, and thus, the states described will be CV cluster states. 

\subsection{Preliminaries}
Before constructing a model for one-way quantum computing using the 3D cluster state, we requires some additional definitions.

The first step is to define the relevant modes that the CV cluster state is made from in terms of the infinitesimal spectral modes described in the previous section. These can be written as an integral over the squeezing spectrum weighted by a normalized Kernel function, $K(\omega)$, within the frequency range $\left[-\Delta\omega/2,\Delta\omega/2\right]$, with time index $t_{n}$, spectral index $l$ and a spatial mode index $s$. We collect all the field operators $\delta\hat{a}_{\text{s},\,l,\,t_{n}}(\omega)$, into a vector $\hat{\mathbf{a}}$:
\begin{equation}
\begin{aligned}
    \hat{\mathbf{a}}&\equiv\bigoplus_{\substack{s\in\left\{\text{a,b,c,d}\right\}\\l\leq l_{\rm 3dB}\\t_{n}\in T}}\int_{-\frac{\Delta\omega}{2}}^{\frac{\Delta\omega}{2}}K(\omega)\;\delta\hat{a}_{\text{s},\,l,\,t_{n}}(\omega)\mathrm{d}\omega\\
    &=\left(\hat{a}_{1},\hat{a}_{2},\hat{a}_{3},\hat{a}_{4},\cdots,\hat{a}_{n},\cdots\right)^{\text{T}}. \label{eq:modesdef}
\end{aligned}
\end{equation}
The subscript of $\hat{a}_{j}$ denotes the modes on a particular graph representation of CV cluster state. The entanglement was previously characterized by spatial, spectral, and temporal modes, but now is only characterized by a single subscript $j$. The length of $\hat{\mathbf{a}}$ is set by the number of entangled modes in Eq.~(\ref{EQ:SIZE}). 

Relative to these operators, recall that the quadrature operators can be defined via $\hat{a}_{j}=(\hat{q}_{j}+i\hat{p}_{j})/\sqrt{2}$. We denote the $s^{\text{th}}$ basis state of the operator $\hat{r}_{j}$ as $\ket{s}_{r_{l}}$, where $\hat{r}$ will usually be a position or momentum operator, or a linear combination of the two. 

Now we define some useful gates to construct our measurement-based protocol. 
The 50:50 beamsplitter gate (BSG) between modes $(j\,,k)$ can be written as
\begin{equation}
        \hat{B}_{jk}\equiv e^{-\frac{\pi}{4}\left(\hat{a}_{j}^{\dagger} \hat{a}_{k}-\hat{a}_{k}^{\dagger} \hat{a}_{j}\right)}.
\end{equation}
Note that this gate is not invariant under a swapping of the inputs. Graphically, this is represented by a red arrow from mode $j$ to mode $k$.

A specific combination of these results in a balanced mixing of four modes, which we call the \emph{foursplitter} gate
\begin{align} 
\hat{A}_{jklm} \equiv \hat{B}_{jk}\hat{B}_{lm}\hat{B}_{jl}\hat{B}_{km}. \label{EQ:FSdef}
\end{align}

It will become convenient to use the matrix representation of the Heisenberg-picture evolution for these gates, i.e., 
\begin{align}
\hat{B}^{\dagger}_{jk} \begin{pmatrix} \hat{a}_{j} \\ \hat{a}_{k}\end{pmatrix} \hat{B}_{jk} & = \mathbf{B}  \begin{pmatrix} \hat{a}_{j} \\ \hat{a}_{k}\end{pmatrix},
\label{EQ:BSmat}
\end{align}
and
\begin{align}
\hat{A}^{\dagger}_{jklm} \begin{pmatrix} \hat{a}_{j} \\ \hat{a}_{k}\\ \hat{a}_{l}\\ \hat{a}_{m}\end{pmatrix} \hat{A}_{jklm} & = \mathbf{A}  \begin{pmatrix} \hat{a}_{j} \\ \hat{a}_{k} \\ \hat{a}_{l} \\ \hat{a}_{m} \end{pmatrix},
\end{align}
where
\begin{equation}
    \mathbf{B}=\frac{1}{\sqrt{2}}\begin{pmatrix} 1 & -1 \\ 1 & 1 \end{pmatrix}\;\;,\;\;
    \mathbf{A}=\frac{1}{2}
    \begin{pmatrix} 1 & -1 & -1 & 1 \\ 1 & 1 & -1 & -1 \\ 1 & -1 & 1 & -1 \\ 1 & 1 & 1 & 1 \end{pmatrix}.
\end{equation}

The phase delay is written as 
\begin{align}
        \hat{R}(\theta)\equiv e^{i\theta\,\hat{a}^{\dagger}\hat{a}}.
\end{align}
The single mode squeezer is given the following nonstandard definition:
\begin{equation}
     \hat{S}(s)\equiv\hat{R}(\text{Im}\ln s) \exp\left[ -\frac{1}{2} \left(\text{Re} \ln s\right)\left(\hat{a}^{2} - \hat{a}^{\dagger 2}\right)\right],
\end{equation}
where $s$ is known as the \emph{squeezing factor}, which is the ordinary squeezing gate with \emph{squeezing paramater} $r=\ln|s|$ followed by a $\pi$ phase delay if $s<0$.

The displacement operator is defined as
\begin{equation}
    \hat{D}(\alpha)\equiv e^{\alpha \hat{a}^{\dagger}-\alpha^*\hat{a}}.
\end{equation}
Finally, it will be convenient to define the following single-mode Gaussian unitary
\begin{equation}
    \hat{V}\left(\theta_{j}, \theta_{k}\right)\equiv \hat{R}\left(\frac{\theta_{j}+\theta_{k}}{2}\right) \hat{S}\left(\tan \frac{\theta_{j} -\theta_{k}}{2} \right)\hat{R}\left(\frac{\theta_{j} +\theta_{k}}{2}\right).
\end{equation}

\subsubsection{Square cluster state}
The following square cluster state plays a key role in the analysis of our one-way quantum-computing protocol. It can be generated by sending one mode from each of a pair of two-mode CV cluster states through a 50:50 BSG
\begin{align}
    \includegraphics[width=0.9\linewidth]{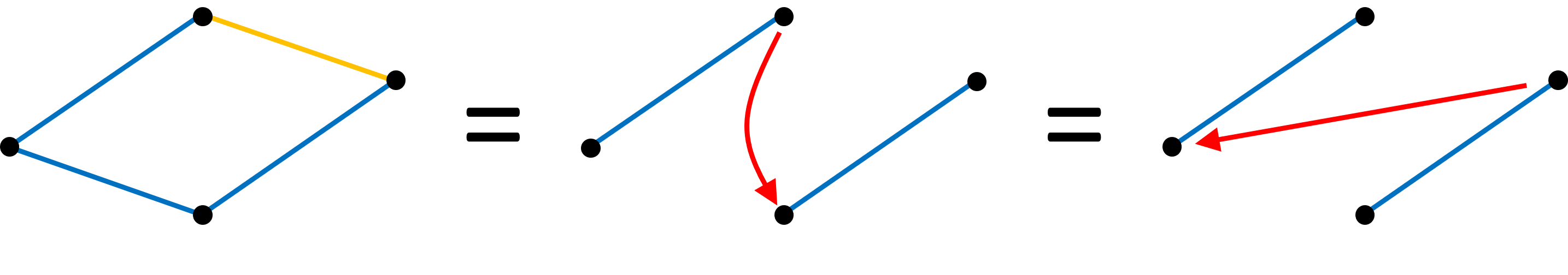}\label{EQ:p2sq}
\end{align}
where $C=1/\sqrt{2}$ on the left and $C=1$ on the center and right.

\subsubsection{Physical modes and distributed modes}

As with other multilayered CV cluster states~\cite{Alexander14, Alexander16, Alexander16a}, the description of one-way quantum computing can be simplified by expressing it in terms of so-called \emph{distributed modes}, which defines a nonlocal tensor product structure for each macronode. Each of the physical modes $\left\{\text{a, b, c, d}\right\}$ within a given macronode is mapped to a distinct \emph{distributed mode} in $\left\{\alpha, \beta, \gamma, \delta\right\}$ via
\begin{align}
\begin{pmatrix}
\hat{a}_{\alpha}\\
\hat{a}_{\beta}\\
\hat{a}_{\gamma}\\
\hat{a}_{\delta}
\end{pmatrix} &\equiv
\mathbf{A}^{-1}
\begin{pmatrix}
\hat{a}_{\text{a}}\\
\hat{a}_{\text{b}}\\
\hat{a}_{\text{c}}\\
\hat{a}_{\text{d}}
\end{pmatrix}. \label{EQ:distfromphys}
\end{align}

\begin{figure*}
    \centering
    \includegraphics[width=0.8\linewidth]{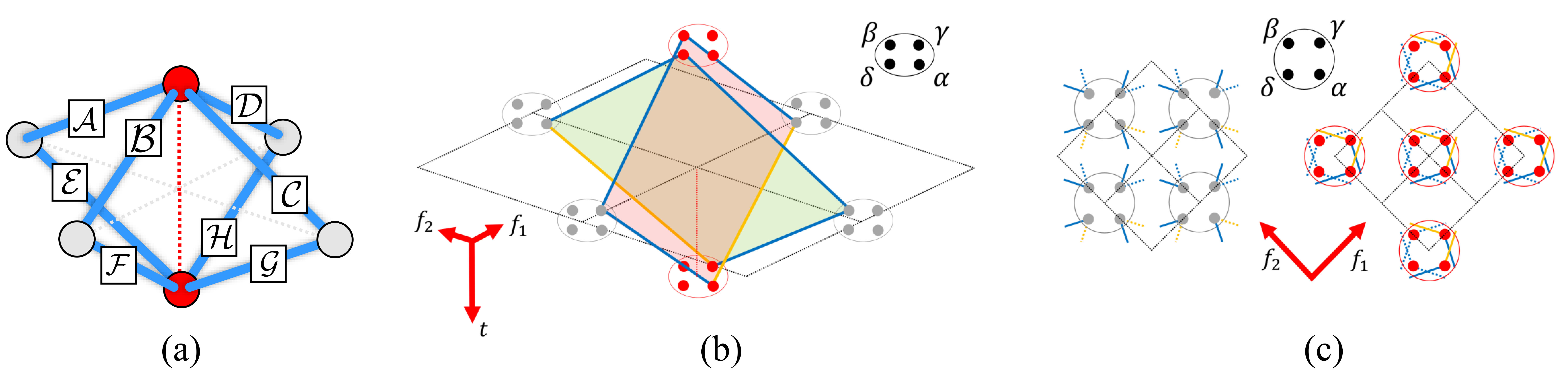}
    \caption{(a) Unit cell of the 3D cluster state with respect to physical modes. Every macronode is connected to eight other macronodes. Furthermore, each individual mode has $8\times4=32$ neighbours. Edge conventions are as defined in Fig.~\ref{FIG:3DCVcsgraph2}.  
   (b) Unit cell of the 3D cluster state graph in terms of distributed modes `$\alpha$', `$\beta$', `$\gamma$' and `$\delta$' (see Eq.~(\ref{EQ:distfromphys})). (c)  Top view of (b). Constant time cross section showing a layer of grey and red macronodes and their connections to macronodes in the layers above and below. Solid (dotted) lines represent upwards (downwards) pointing edges.}
    \label{FIG:logpic}
\end{figure*}

Expressing the 3D cluster state in terms of the distributed modes simplifies the graph substantially. It becomes a disjoint collection of square cluster states as shown in Fig.~\ref{FIG:logpic}.

In our protocol we encode input states into half of the macronodes of a given time step---specifically the red ones shown in Fig.~\ref{FIG:3DCVcsgraph2}. The gray macronodes serve as ``{\em routers}'' that control the application of entangling gates as the inputs distribute through the cluster state in the time direction. Besides simplifying the graph, the distributed modes play a special role in defining how each input is encoded within a given red macronode on a given time slice. More concretely, we will choose to encode each input into either the `$\alpha$' or the `$\gamma$' distributed mode. In fact, it will be convenient to change whether the input resides in either the `$\alpha$' or `$\gamma$' mode from time step to time step.   

\begin{figure*}
    \centering
    \includegraphics[width=1\linewidth]{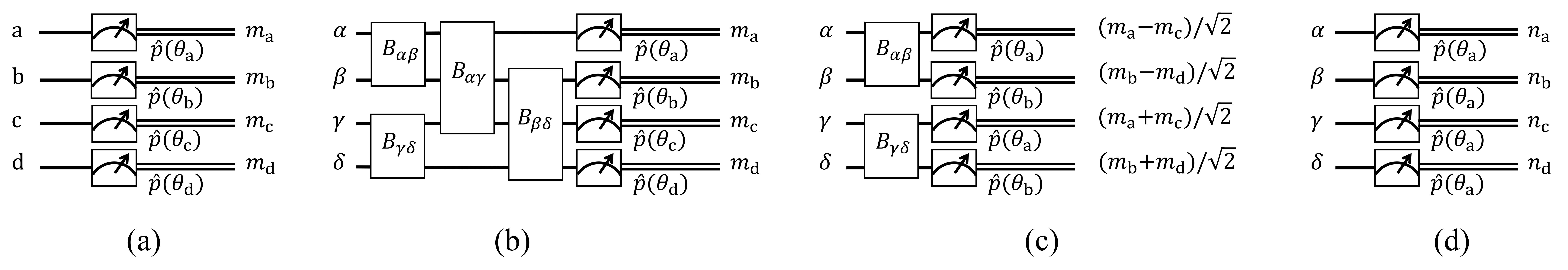}
    \caption{(a) Local measurement of the physical modes via Eq.~(\ref{eq:pmeas}). (b) From the perspective of the distributed modes, such a measurement will in the general case project onto entangled states. (c) For some choices of angles, these projections will be onto partially or fully separable states. Choosing ${\theta_{\text{a}}=\theta_{\text{c}}}$ and ${\theta_{\text{b}}=\theta_{\text{d}}}$ makes the measurement separable with respect to the {$\left(\alpha, \beta\, \vert\, \gamma, \delta\right)$} bipartition. This follows from two applications of Eq.~(\ref{EQ:BSpropmeas}). (d) Choosing $\theta_{\text{a}}=\theta_{\text{b}}=\theta_{\text{c}}=\theta_{\text{d}}$ results in a local measurement with respect to the distributed modes. We define $(n_{\text{a}}, n_{\text{b}}, n_{\text{c}}, n_{\text{d}})^{\text{T}}=\mathbf{A} (m_{\text{a}}, m_{\text{b}}, m_{\text{c}}, m_{\text{d}})^{\text{T}}$. This follows from four applications of Eq.~(\ref{EQ:BSpropmeas}).}
    \label{FIG:distmeas}
\end{figure*}

Our procotol for one-way quantum computing involves local homodyne measurements with respect to the physical modes. We denote the measured bases as 
\begin{align}
    \hat{p}_{j}(\theta)\equiv \cos{\theta}\,\hat{p}_{j} + \sin{\theta}\,\hat{q}_{j}, \label{eq:pmeas}
\end{align}
where the angle $\theta$ is controlled by the relative phase between physical mode $j$ and its corresponding local oscillator.
 Local measurements on the {\em physical} modes (as shown in Fig.~\ref{FIG:distmeas} (a)) translate to quantum gates followed by local measurements on the {\em distributed} modes (as shown in Fig.~\ref{FIG:distmeas} (b)). However, for special choices of the measurement angles, this measurement can appear as partially separable (see Fig.~\ref{FIG:distmeas} (c)) or completely separable (see Fig.~\ref{FIG:distmeas} (d)). This follows from the following identity in Ref.~\cite{Alexander16}:
\begin{align}
    \bra{m_{j}}_{p_{j}(\theta)}\bra{m_{k}}_{p_{k}(\theta)}\hat{B}_{jk} = \left\langle \frac{m_{j} -m_{\text{k}}}{\sqrt{2}}\right\vert_{p_{j}(\theta)} \left\langle \frac{m_{j} +m_{k}}{\sqrt{2}}\right\vert_{p_{k}(\theta)}. \label{EQ:BSpropmeas}
\end{align}

It will also be useful to note that introducing an arbitrary permutation on the four modes before measurement shown in Fig.~\ref{FIG:distmeas} (b) is equivalent to swapping some of the measurement bases after the four splitter and changing the sign of some outcomes~\cite{Alexander16a}. Let $\hat{P}$ be an operator that permutes modes $\{\alpha, \beta, \gamma, \delta\}$. Then 
\begin{align}
    \hat{P}^{\dagger} \hat{A}^{\dagger}_{\alpha\beta\gamma\delta}
    \begin{pmatrix}
    \hat{p}_{\alpha}(\theta_{\text{a}}) \\ 
      \hat{p}_{\beta}(\theta_{\text{b}}) \\
        \hat{p}_{\gamma}(\theta_{\text{c}}) \\
          \hat{p}_{\delta}(\theta_{\text{d}}) 
    \end{pmatrix} \hat{A}_{\alpha\beta\gamma\delta} \hat{P} = 
     \hat{A}^{\dagger}_{\alpha\beta\gamma\delta}
    \mathbf{M}\begin{pmatrix}
    \hat{p}_{\alpha}(\theta_{\text{a}}) \\ 
      \hat{p}_{\beta}(\theta_{\text{b}}) \\
        \hat{p}_{\gamma}(\theta_{\text{c}}) \\
          \hat{p}_{\delta}(\theta_{\text{d}}) 
    \end{pmatrix} \hat{A}_{\alpha\beta\gamma\delta}, \label{EQ:permmeas}
\end{align}
where the left-hand side is the Heisenberg picture evolution of the vector of observables measured in the circuit from Fig.~\ref{FIG:distmeas} (b) through the permutation gate $\hat{P}$. The right-hand side shows that this is equivalent to multiplying the vector of observables on the right-hand side by a $4\times 4$ matrix $\mathbf{M}$, which is the product of a permutation matrix and a diagonal matrix with entries in $\{ -1, 1\}$~\cite{Alexander16a}. Therefore, introducing a permutation gate before such a measurement is equivalent to swapping the measurement bases at the homodyne detectors in   Fig.~\ref{FIG:distmeas} (a), and flipping the sign of some outcomes.

\subsubsection{Quantum computing via teleportation}
Given an input, a two-mode CV cluster state, and a measurement that implements a 50:50 BSG followed by local homodyne detection, i.e., 
\begin{align}
    \includegraphics[width=0.5\linewidth]{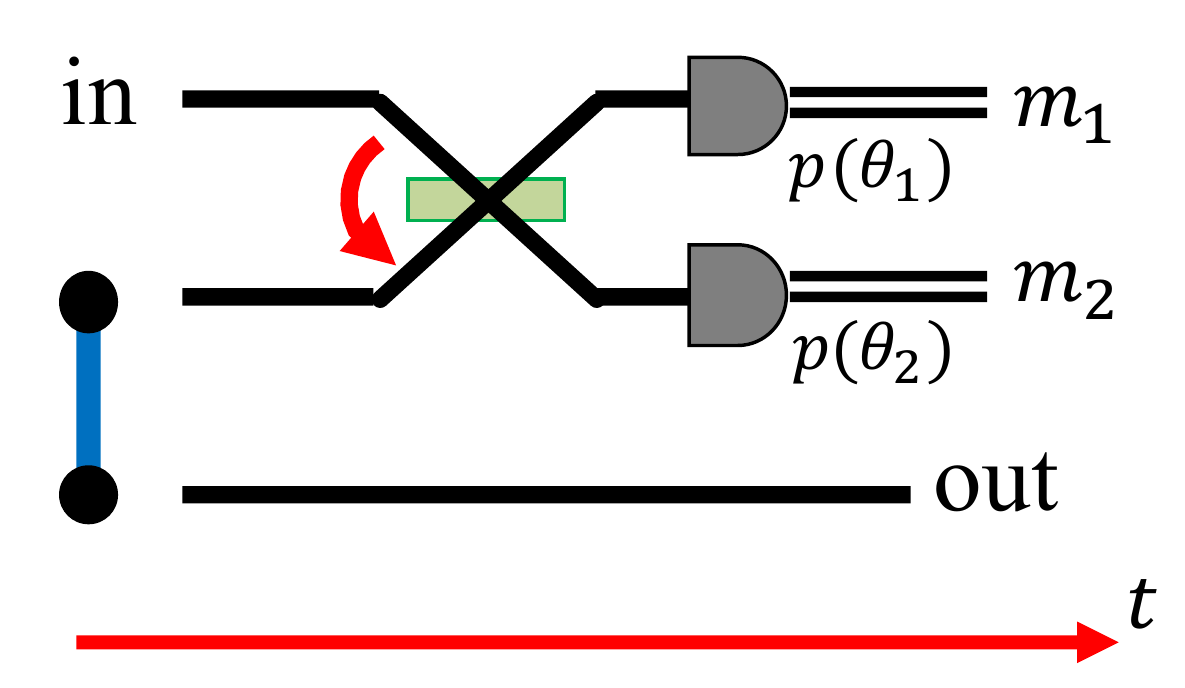}\label{EQ:tel}
\end{align}
we can implement the gate 
\begin{align}
    \hat{D}\left[\frac{-ie^{i\theta_{2}}m_{1} - i e^{i\theta_{1}}m_{2}}{\sin(\theta_{1}-\theta_{2})}\right]
    \hat{V}\left(\theta_{1}, \theta_{2}\right)
\end{align}
via teleportation, where we have assumed infinite squeezing~\cite{Alexander18, Alexander14}. Finite squeezing effects for this gate can be included via the analysis described in Ref.~\cite{Alexander14}. 
The teleportation-induced gate $\hat{V}$ is a crucial factor for the one-way quantum-computing protocol described below. For convenience, we assume all measurement outcomes equal to zero. The true evolution will only differ from this case by a final displacement since all gates described are Gaussian.

\begin{figure}
	{\centering\includegraphics[width=1\linewidth]{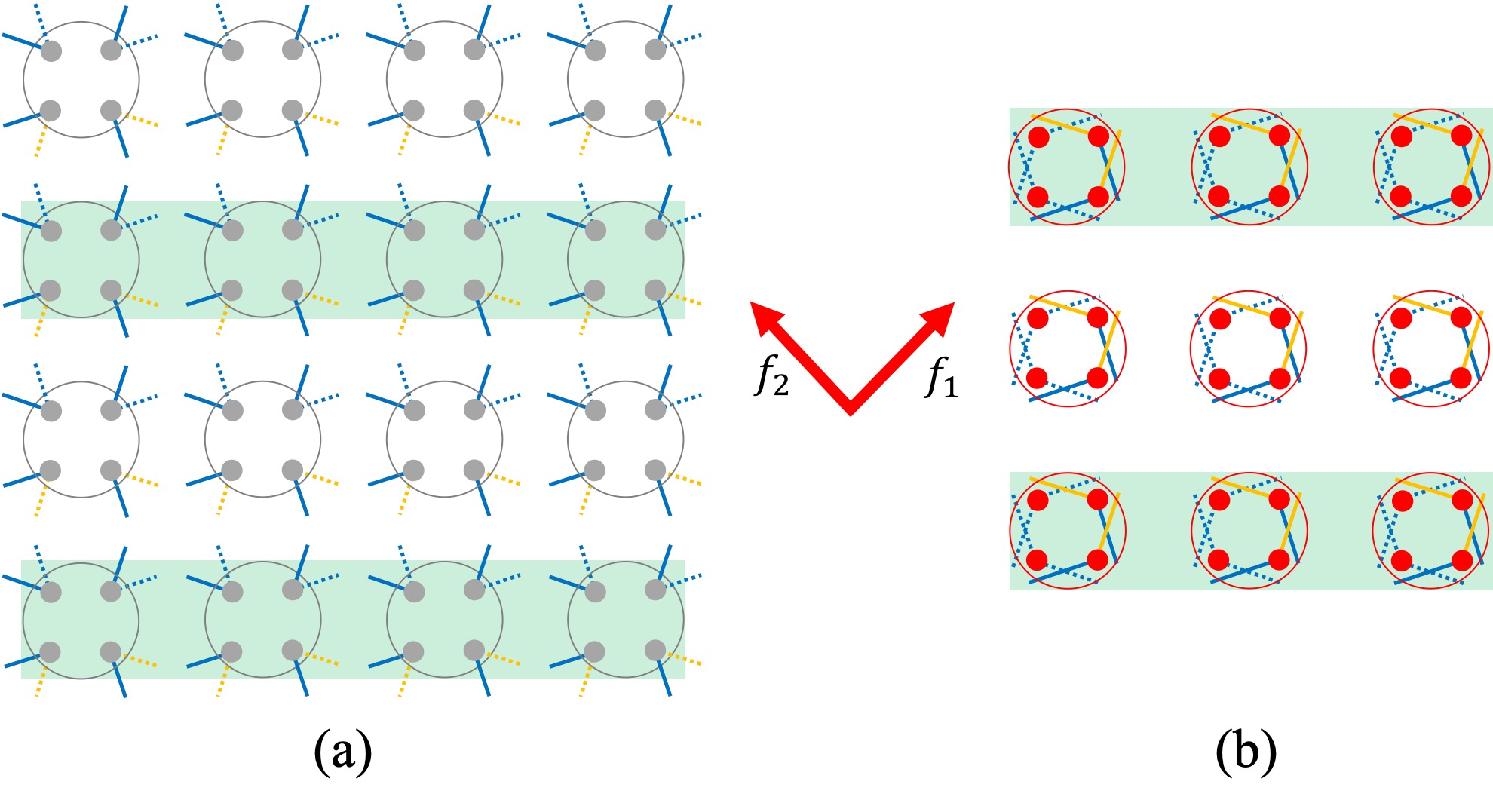}\\}
	\caption{\label{FIG:qpmeas} (a) In order to implement single-mode gates on all inputs, the grey macronodes in the (un-)shaded regions are measured in the $\hat{p}(\pm\pi/4)$ basis, respectively. (b) The relative sign in the gate implemented in Eq.~(\ref{EQ:smgate}) depends on whether the input lies in a green shaded region, or not. Diagrammatic conventions are the same as in Fig.~\ref{FIG:logpic} (b) and (c).}
\end{figure}

\begin{figure}
    \includegraphics[width=1\linewidth]{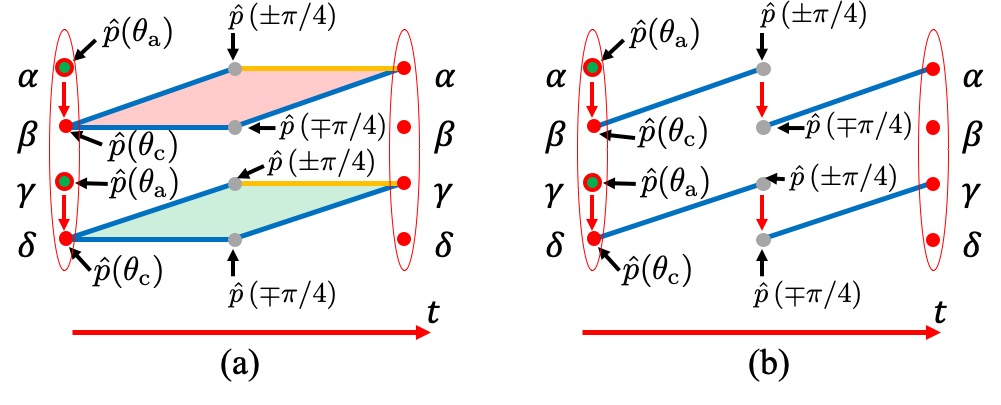}
    \caption{\label{FIG:drw}
 (a) The two square graphs from Fig.~\ref{FIG:logpic} have been redrawn side by side. Note that the time arrow is from left to right, rather than from top to bottom. The green modes on the left are the distributed modes that can contain an input. The red arrows denote the application of a 50:50 BSG, and the black arrows indicate the local basis measurement after all such BSGs. (b) Alternative representation of (a). The square graph cluster state is equivalent to two pairs stitched together by a 50:50 BSG (see Eq.~(\ref{EQ:p2sq})). This picture more clearly shows that the measurement pattern implements sequential teleportation (see Eq.~(\ref{EQ:tel})). }
\end{figure}

Now, we consider performing measurements on the top red and four grey macronodes in Fig.~\ref{FIG:logpic} (b). Each macronode is measured as shown in Fig.~\ref{FIG:distmeas} (a) and (b).  For the red macronode, we set {$\theta_{\text{a}}=\theta_{\text{c}}$}, and {$\theta_{\text{b}}=\theta_{\text{d}}$}, and the measurement can be modelled as shown in Fig.~\ref{FIG:distmeas} (c). Subsequently, we choose to measure all grey macronodes in either the $\hat{p}(\pm \pi/4)$ basis, where the sign of the angle is determined with respect to Fig.~\ref{FIG:qpmeas} (a). Since each mode within a given grey macronode is measured in the same basis, measurement of the physical modes can be modelled as shown in Fig.~\ref{FIG:distmeas} (d). The overall measurement pattern can now be summarized as shown in Fig.~\ref{FIG:drw} (a).

Equivalently, we could use the description in Fig.~\ref{FIG:drw} (b), which shows more clearly that this measurement pattern implements two rounds of teleportation (see Eq.~(\ref{EQ:tel})). Thus, we can write the gate implemented as  
\begin{align}
\hat{V}_{\alpha} \left(\pm\frac{\pi}{4},\mp\frac{\pi}{4}\right)\hat{V}_{\gamma} \left(\pm\frac{\pi}{4},\mp\frac{\pi}{4}\right) \hat{V}_{\alpha} (\theta_{\text{a}}, \theta_{\text{b}})\hat{V}_{\gamma}(\theta_{\text{a}}, \theta_{\text{b}}) , \label{EQ:smgate}
\end{align}
where the $\pm$ sign depends on whether the red macronode is within a green shaded region or an unshaded region in Fig.~\ref{FIG:qpmeas}~(b), respectively. Note that after teleportation, the input resides in the bottom red macronode of Fig.~\ref{FIG:logpic} (b). 

Evolving under Eq.~(\ref{EQ:smgate}) is equivalent to using two decoupled copies of the \emph{dual rail wire}~\cite{Alexander14}. Two successive rounds of this measurement pattern are sufficient to implement arbitrary single-mode Gaussian unitary gates for inputs in either the `$\alpha$' and `$\gamma$' distributed modes~\cite{Ukai10, Alexander14}.

Another important measurement-based operation for our one-way quantum computing protocol is a swap between modes `$\alpha$' and `$\gamma$', thus changing the distributed mode in which the logical information resides. As described in Eq.~(\ref{EQ:permmeas}), a swap before a macronode measurement is equivalent to permuting the homodyne angles and post-processing.  By swapping between the `$\alpha$' and `$\gamma$' distributed modes, one can use the BMZI in Fig.~\ref{FIG:Programmable} to insert an input state, such as a GKP ancilla state, into either distributed mode of any macronode in the cluster state.

\subsubsection{Entangling gates}
\begin{figure*}
    \centering
    \includegraphics[width=0.93\linewidth]{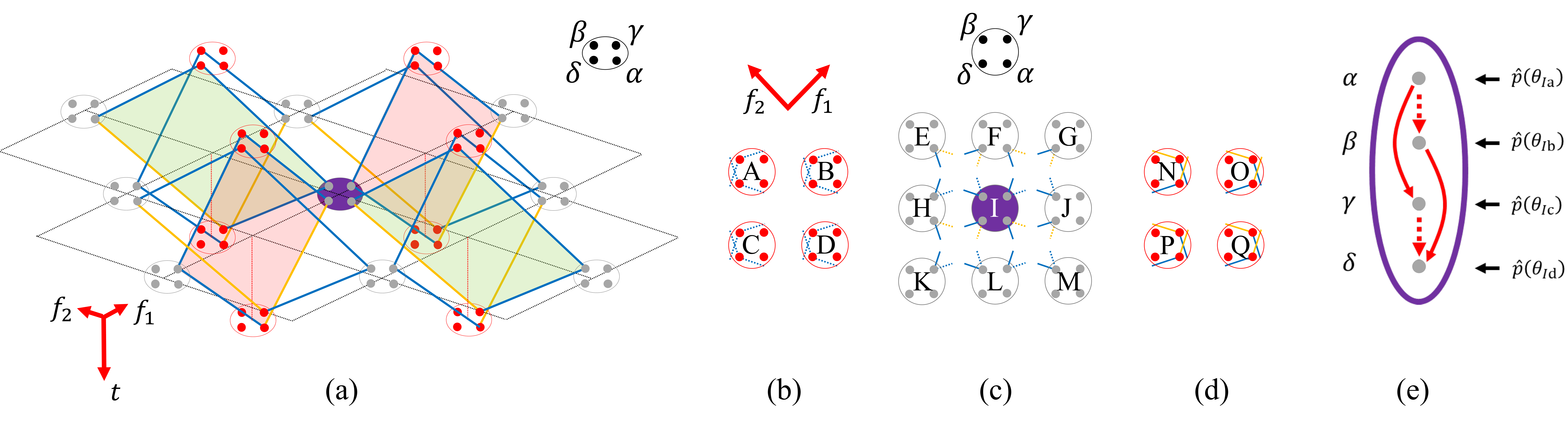}
    \caption{ (a) Subgraph of the full 3D cluster state represented using distributed modes. Some of the square graphs are shaded red or green to make the 3D layout clearer. (b) The top layer of the graph viewed from above. Macronodes `A', `B', `C', and `D'  contain input states encoded within `$\alpha$' and `$\gamma$' distributed modes. Solid lines are pointing out of the page, while dotted lines are pointing into the page. (c) Middle layer of the graph viewed from above.  (d) Bottom layer of the graph viewed from above. These modes are not measured in this round, but will contain input states after the upper two layers are measured. (e) Macronode `I' is the only grey macronode where the homodyne angles are not all the same. The dotted arrows represent 50:50 BSGs that act before those represented by the solid arrows. After all 50:50 BSGs, the modes in this macronode are measured in the bases shown in black.}
    \label{FIG:multilog}
\end{figure*}

Next we describe how to implement multimode (a.k.a. entangling) gates between inputs encoded within adjacent red macronodes. We consider the four red macronodes adjacent to a particular grey macronode. In order to perform an entangling operation between any subset of the inputs on these four red macronodes, the only change relative to the single-mode gates described in the previous section is that modes that make up the central grey macronode are measured in different bases. For concreteness, we consider a particular subgraph of the 3D cluster state, as shown in Fig.~\ref{FIG:multilog}.

We will assume that if a grey macronode is used to implement entangling gates, then none of the adjacent grey macronodes are used to do so as well. With respect to Fig.~\ref{FIG:multilog}~(c), this means that modes in macronodes `E', `F', `G', `K', `L', and `M' are all measured in the $\hat{p}(\pm\pi/4)$ basis, and modes in macronodes `H' and `J' are measured in the $\hat{p}(\mp\pi/4)$ basis, as described in the previous section. 
In order for inputs in macronodes `A' and `D' (`B' and `C') to participate in the entangling gate, they will be assumed to reside in the `$\gamma$' (`$\alpha$') distributed mode. Note that if the input happened to be in the other of the two distributed modes, a swap can be employed as described in the previous section.

Denote the measured bases for the red macronodes $R \in\{\text{A, B, C, D}\}$ as  $\hat{p}_{R\text{a}}(\theta_{R\text{a}}), \hat{p}_{R\text{b}}(\theta_{R\text{b}}), \hat{p}_{R\text{c}}(\theta_{R\text{a}})$ and $\hat{p}_{R\text{d}}(\theta_{R\text{b}})$. Similarly, denote the measured bases for macronode `I' (shaded purple in Fig.~\ref{FIG:multilog}) as $\hat{p}_{\text{Ia}}(\theta_{\text{Ia}}), \hat{p}_{\text{Ib}}(\theta_{\text{Ib}}), \hat{p}_{\text{Ic}}(\theta_{\text{Ic}})$ and $\hat{p}_{\text{Id}}(\theta_{\text{Id}})$.

Excepting special cases such as those mentioned in Fig.~\ref{FIG:distmeas} (b) and (c), generic angles $\theta_{\text{Ia}}, \theta_{\text{Ib}}, \theta_{\text{Ic}}$, and  $\theta_{\text{Id}}$ will result in measurements as shown in Fig.~\ref{FIG:distmeas} (b). The 50:50 BSGs acting before the measurement device in Fig.~\ref{FIG:distmeas} (b) are represented  graphically in Fig.~\ref{FIG:multilog}~(e). By employing a series of identities for 50:50 BSGs acting on entangled pairs, we can move the BSGs so that they act on other modes, thereby reducing the measurement on macronode `I' to one that is local with respect to the distributed mode tensor product structure. This technique is known as \emph{beamsplitter gymnastics}~\cite{Alexander16}.

\begin{figure*}
    \centering
    \includegraphics[width=\linewidth]{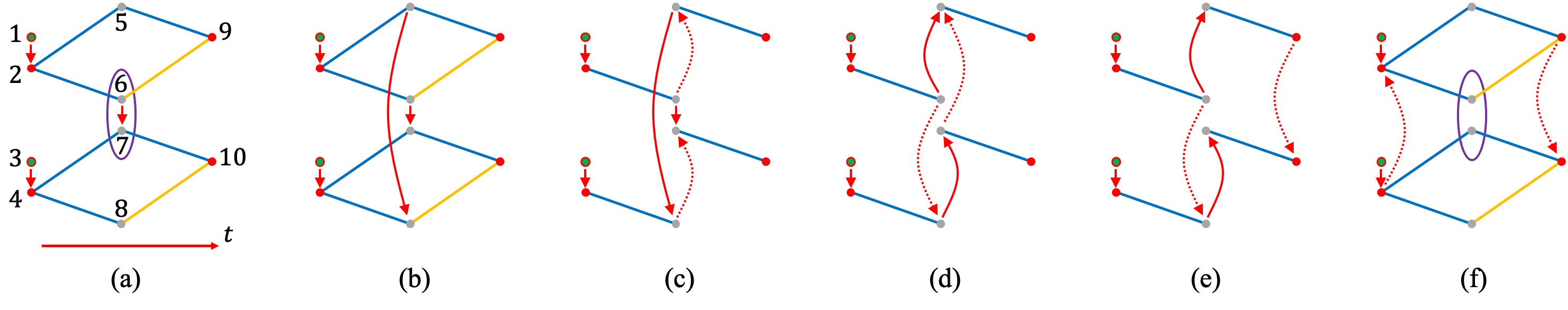}
    \caption{ Beamsplitter gymnastics for $\hat{B}_{\text{I}\alpha,  \text{I}\beta}$ and $\hat{B}_{\text{I}\alpha, \text{I}\gamma}$. For $\hat{B}_{\text{I}\alpha,  \text{I}\beta}$, modes $(1-10) = (\text{D}\gamma, \text{D}\delta, \text{A}\gamma, \text{A}\delta, \text{M}\beta, \text{I}\alpha, \text{I}\beta, \text{E}\alpha, \text{Q}\gamma, \text{N}\gamma)$. For $\hat{B}_{\text{I}\alpha, \text{I}\gamma}$, modes $(1-10)= (\text{D}\gamma, \text{D}\delta, \text{A}\gamma, \text{A}\delta, \text{M}\beta, \text{I}\alpha, \text{I}\gamma, \text{G}\delta, \text{Q}\gamma, \text{O}\alpha)$. (a) The goal is to move the 50:50 BSG between the grey modes in the purple oval (6 and 7), replacing it with an operation that only acts on different modes. (b) Since modes 5 and 8 are measured in the same basis, we can use postprocessing to insert an extra 50:50 BSG (see Eq.~(\ref{EQ:BSpropmeas})). (c) Each square graph can be replaced with two entangled pairs and a pair of 50:50 BSG. The dotted lines indicate that these act \emph{before} the other 50:50 BSGs (see Eq.~(\ref{EQ:p2sq})). (d) By direct calculation using Eqs.~(\ref{EQ:BSmat}),  $\hat{B}_{5,8}\hat{B}_{6,7}\hat{B}_{6,5}\hat{B}_{8,7}=\hat{B}_{6,5}\hat{B}_{8,7}\hat{B}_{6,8}\hat{B}_{7,5}$. (e) We can move the rightmost dotted beamsplitter using the second equality in Eq.~(\ref{EQ:p2sq}). (f) Similarly, the leftmost dotted beamsplitter can be moved to the left and we can substitute the square graphs back in by four applications of Eq.~(\ref{EQ:p2sq}). At this stage, the 50:50 BSG $\hat{B}_{6,7}$ had been replaced with the two dotted 50:50 BSGs that act on either sides of the square graphs, thereby achieving our goal.}
    \label{FIG:BSgym}
\end{figure*}

\begin{figure}
    \centering
    \includegraphics[width=0.9\linewidth]{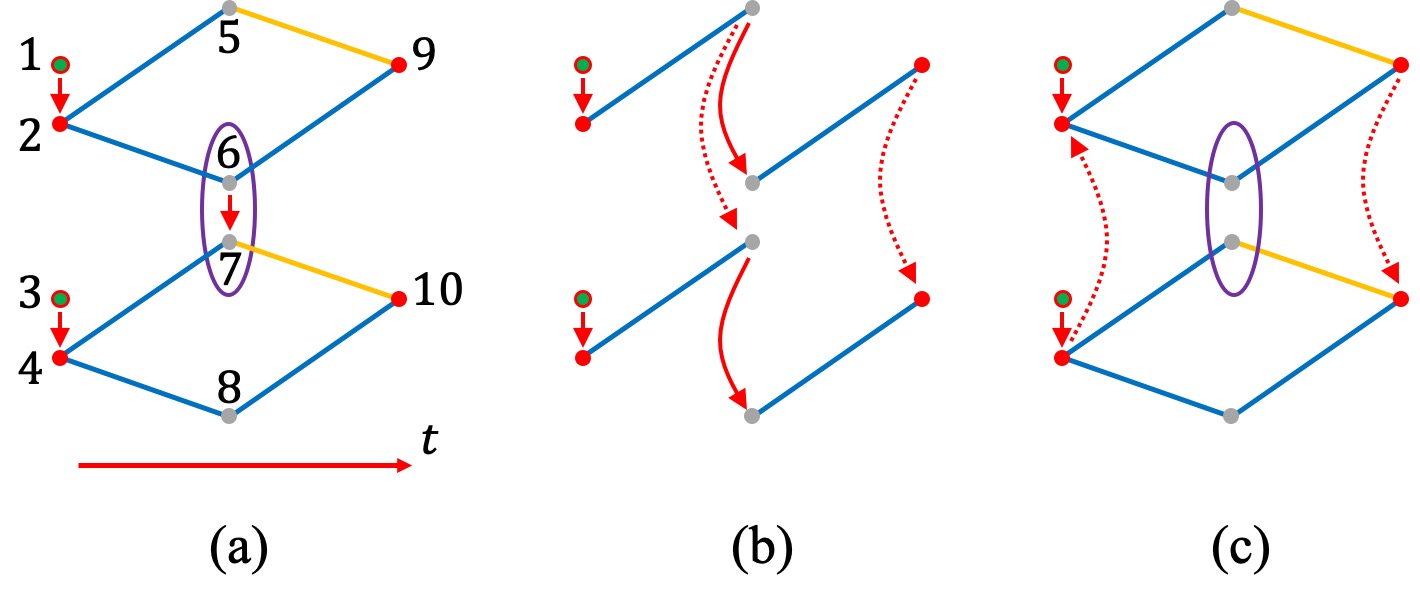}
    \caption{ Beamsplitter gymnastics for $\hat{B}_{\text{I}\gamma, \text{I}\delta}$ and $\hat{B}_{\text{I}\beta, \text{I}\delta}$.  For $\hat{B}_{\text{I}\gamma, \text{I}\delta}$, modes ${(1-10)} = (\text{B}\alpha, \text{B}\beta, \text{C}\alpha, \text{C}\beta, \text{G}\delta, \text{I}\gamma, \text{I}\delta, \text{K}\gamma, \text{O}\alpha, \text{C}\alpha)$. For $\hat{B}_{\text{Ib, Id}}$, modes ${(1-10)}= (\text{A}\gamma, \text{A}\delta, \text{C}\alpha, \text{C}\beta, \text{E}\alpha, \text{I}\beta, \text{I}\delta, \text{K}\gamma, \text{N}\gamma, \text{P}\alpha)$. (a) The goal is to move the BSG between the grey modes in the purple oval (6 and 7), replacing it with an operation that only on other modes. (b) By following a similar sequence of steps as in Fig.~\ref{FIG:BSgym} (or equivalently, using the proof shown in Ref.~\cite{Alexander16}) we arrive at this alternative description. (c) The leftmost dotted beamsplitter can be moved to the left and we can substitute the square graphs back in by four applications of Eq.~(\ref{EQ:p2sq}). At this stage, the 50:50 BSG $\hat{B}_{6,7}$ had been replaced with the two dotted 50:50 BSGs that act on either sides of the square graphs, thereby achieving our goal. }
    \label{FIG:BSgym2}
\end{figure}

Fig.~\ref{FIG:BSgym} shows how to do this for $\hat{B}_{\text{I}\alpha, \text{I}\beta}$ and $\hat{B}_{\text{I}\alpha, \text{I}\gamma}$, and similarly,  Fig.~\ref{FIG:BSgym2} shows how to do this for $\hat{B}_{\text{I}\gamma, \text{I}\delta}$ and $\hat{B}_{\text{I}\beta, \text{I}\delta}$. The beamsplitter positioning in Fig.~\ref{FIG:BSgym}(e) and Fig.~\ref{FIG:BSgym2}(b) are such thet all beamsplitters lie between teleportation steps, and thus, the evolution is merely a sequence of logical BSGs and teleportation. The total evolution can be written as
\begin{align}
   \hat{W}_{A\alpha, B\gamma, C\gamma, D\alpha}(\boldsymbol{\theta})=  &
     \hat{A}_{\text{D}\gamma, \text{A}\gamma, \text{B}\alpha,  \text{C}\alpha} \hat{V}_{\text{A}\alpha}\left(\pm\frac{\pi}{4}, \theta_{\text{I}\text{b}} \right) \hat{V}_{\text{B}\gamma}\left(\pm\frac{\pi}{4}, \theta_{\text{I}\text{c}}\right)  \nonumber \\
    &\hspace{-1cm}\times \hat{V}_{\text{C}\gamma} \left( \theta_{\text{I}\text{d}}, \pm\frac{\pi}{4}\right) \hat{V}_{\text{D}\alpha}\left(\theta_{\text{I}\text{a}}, \pm\frac{\pi}{4}\right) \hat{A}_{\text{D}\gamma,  \text{A}\gamma, \text{B}\alpha,  \text{C}\alpha} \nonumber \\
      &\hspace{-1cm} \times\left[\prod_{R\in\{\text{A}\gamma, \text{B}\alpha, \text{C}\alpha, \text{D}\gamma\}}\hat{V}_{R}(\theta_{R\text{a}}, \theta_{R\text{b}})\right].
    \label{EQ:4Ugen}
\end{align}
Any inputs present in the alternative distributed modes $j\in\left\{ \text{A}\alpha, \text{B}\gamma, \text{C}\gamma, \text{D}\alpha \right\}$ evolve according to the single-mode protocol described in the previous section
\begin{align}
     \hat{V}_{j}\left(\pm\frac{\pi}{4}, \mp\frac{\pi}{4}\right)
    \hat{V}_{j}(\theta_{j\text{a}}, \theta_{j\text{b}}).
    \label{EQ:4Ugensm}
\end{align}
Therefore, by the applying the swap degree of freedom between `$\alpha$' and `$\gamma$' distributed modes, we can control participation in the entangling gate.

The four-mode entangling gate $\hat{W}$ can be simplified for particular choices of homodyne angles. Below we describe various restrictions on the angles that result in gates that entangling gates between any two pairs of modes within $\{\text{A}\gamma, \text{B}\alpha, \text{C}\alpha, \text{D}\gamma\}$. 
We define
\begin{align}
&\hat{G}_{jk}^{\pm}(\theta_{\text{1}},\theta_{\text{2}}, \theta_{\text{3}}, \theta_{\text{4}}, \theta_{\text{5}}, \theta_{\text{6}})\equiv\\
& \hat{B}_{jk}\left[\hat{V}_{j}\left(\pm\frac{\pi}{4}, \theta_{5}\right) \hat{V}_{k}\left(\theta_{6}, \pm\frac{\pi}{4}\right) \right]
\nonumber \hat{B}_{jk} \left[\hat{V}_{j}( \theta_{1}, \theta_{2}) \hat{V}_{k}(\theta_{3}, \theta_{4})\right].
\end{align}

Restricting $\theta_{\text{I}\text{a}} = \theta_{\text{I}\text{b}}$ and $\theta_{\text{I}\text{c}}=\theta_{\text{I}\text{d}}$ simplifies $\hat{W}$ to
\begin{align}
    &\hat{R}_{\text{B}\alpha}\left(\pi\right)\hat{R}_{\text{D}\gamma}(\pi) \hat{G}^{\pm}_{\text{D}\gamma,\text{B}\alpha}(\theta_{\text{D}\text{a}}, \theta_{\text{D}\text{b}}, \theta_{\text{B}\text{a}}, \theta_{\text{B}\text{b}}, \theta_{\text{I}\text{a}}, \theta_{\text{I}\text{c}}) \nonumber\\
    &\times \hat{G}^{\pm}_{\text{A}\gamma, \text{C}\alpha}(\theta_{\text{A}\text{a}}, \theta_{\text{A}\text{b}}, \theta_{\text{C}\text{a}}, \theta_{\text{C}\text{b}},  \theta_{\text{I}\text{a}}, \theta_{\text{I}\text{c}}),
  \end{align}
which implements a pair of two-mode gates between pairs of modes $(\text{D}\gamma, \text{B}\alpha)$ and $(\text{A}\gamma, \text{C}\alpha)$.

Restricting $\theta_{\text{I}\text{b}} = \theta_{\text{I}\text{d}}$ and $\theta_{\text{I}\text{a}}=\theta_{\text{I}\text{c}}$ simplifies $\hat{W}$ to
\begin{align}
    &\hat{R}_{\text{A}\gamma}\left(\pi\right)\hat{R}_{\text{D}\gamma}\left(\pi\right)\hat{G}^{\pm}_{\text{D}\gamma,\text{A}\gamma}(\theta_{\text{D}\text{a}}, \theta_{\text{D}\text{b}}, \theta_{\text{A}\text{a}}, \theta_{\text{A}\text{b}},  \theta_{\text{I}\text{a}}, \theta_{\text{I}\text{b}})\nonumber\\
    &\times \hat{G}^{\pm}_{\text{B}\alpha,\text{C}\alpha}(\theta_{\text{B}\text{a}}, \theta_{\text{B}\text{b}}, \theta_{\text{C}\text{a}}, \theta_{\text{C}\text{b}}, \theta_{\text{I}\text{a}}, \theta_{\text{I}\text{b}}),
\end{align}
which implements a pair of two-mode gates between pairs of modes $(\text{D}\gamma, \text{A}\gamma)$ and $(\text{B}\alpha, \text{C}\alpha)$.

Restricting $\theta_{\text{I}\text{a}} = \theta_{\text{I}\text{d}}$ and $\theta_{\text{I}\text{b}}=\theta_{\text{I}\text{c}}$ simplifies $\hat{W}$ to
\begin{align}
    &\hat{R}_{\text{A}\gamma}\left(\pi\right)\hat{R}_{\text{C}\alpha}\left(\pi\right)\hat{R}_{\text{D}\gamma}\left(\pi\right)\hat{G}^{\pm}_{\text{D}\gamma,\text{C}\alpha}(\theta_{\text{D}\text{a}}, \theta_{\text{D}\text{b}}, \theta_{\text{C}\text{a}}, \theta_{\text{C}\text{b}},  \theta_{\text{I}\text{a}}, \theta_{\text{I}\text{b}})\nonumber\\
    &\times \hat{G}^{\pm}_{\text{A}\gamma,\text{B}\alpha}(\theta_{\text{A}\text{a}}, \theta_{\text{A}\text{b}}, \theta_{\text{B}\text{a}}, \theta_{\text{B}\text{b}}, \theta_{\text{I}\text{a}}, \theta_{\text{I}\text{b}}) \hat{R}_{\text{C}\alpha}(\pi),
\end{align}
which implements a pair of two-mode gates between pairs of modes $(\text{D}\gamma, \text{C}\alpha)$ and $(\text{A}\gamma, \text{B}\alpha)$.

By applying further restrictions, the gates $\hat{G}$ can be reduced into a more familiar and tunable ${\hat{C}_Z(g) =e^{i g\hat{q}\otimes\hat{q}}}$ gate, $\forall {g\in\field{R}}$, which is a standard entangling gate in continuous-variable quantum computing. In particular
\begin{align}
    \hat{G}^{\pm}_{jk}\left(\mp\frac{\pi}{8}, \pm\frac{3\pi}{8}, \mp\frac{\pi}{8}, \pm\frac{3\pi}{8}, \phi \pm \frac{\pi}{4}, \phi \pm \frac{\pi}{4}\right) \nonumber\\ = \left[ \hat{R}_{j}\left( \mp\frac{3\pi}{4}\right)\otimes \hat{R}_{k}\left( \pm\frac{\pi}{4}\right) \right]\hat{C}_{Zjk}\left(2\cot{\phi}\right),
\end{align}
where we have used the protocol from Ref.~\cite{Alexander16}. Thus, the four mode gate $\hat{W}$ can be reduced to pairs of $\hat{C}_Z$ gates with respect to any pairings of modes $\{\text{A}\alpha, \text{B}\gamma, \text{C}\gamma, \text{D}\alpha \}$. 

By leveraging the swap degrees of freedom at red macronodes as well as the above measurement restrictions at grey macronodes, we have shown that it is possible to implement either two ($\hat{G}$) or four ($\hat{W}$) mode entangling gates between nearest neighbor input states.

\subsection{Universal quantum computing and error correction against finite squeezing effects} \label{sec:uniGKP}
Using finitely squeezed continuous-variable cluster states will result in Gaussian noise~\cite{Alexander14}, the strength of which is set by the available amount of squeezing in the cluster state. This effect can be combated using non-Gaussian quantum error correction, such as a supply of GKP qubits, provided that the squeezing in the CV cluster state is sufficiently high. A 20.5 dB upper bound on the amount of squeezing required was given in Ref.~\cite{Menicucci14}, however, it is likely that this bound can be improved by incorporating quantum error correction strategies compatible with 3D entangled resource states~\cite{Fukui17, Vuillot19, Fukui19, Noh19}. 

A non-Gaussian resource is also required to extend the above one-way quantum-computing protocol---which can only implement Gaussian unitary gates, and hence, is classically simulable---to a universal model.  With the access to a supply of GKP qubits, we can generate all of the necessary ingredients for universal quantum computing with Gaussian operations~\cite{Baragiola19}. In principle, all required Gaussian operations can be implemented using the 3D cluster state with homodyne detection via the gate set described above. Though implementing these gates on the 3D cluster state will additionally introduce Gaussian noise that arises due to having only finite squeezing, this can be corrected by using additional GKP ancilla states injected into the state at regular intervals~\cite{Menicucci14}. 

\subsection{Full Architecture}
\begin{figure*}
	{\centering\includegraphics[width= .75\linewidth]{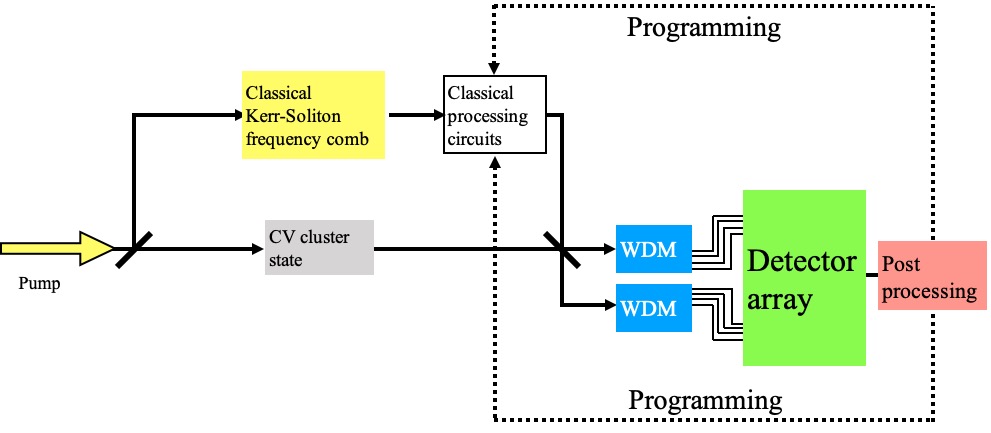}\\}
	\caption{\label{FIG:QC} Flow chart for integrated photonic one-way quantum computing.}
\end{figure*}

All the ingredients introduced above, in conjunction with classical control, yields a universal architecture for quantum computing, as sketched in Fig.~\ref{FIG:QC} and further described below.

The quantum MRs and linear optical components are configured to generate a 3D CV cluster state described in Sec.~\ref{sec:3dCVCSgen}. To implement Gaussian quantum gates, homodyne measurements (with fully tunable and independent local oscillator phases) are performed simultaneously on all spectral modes and sequentially on all temporal modes. To do so, a classical Kerr-Soliton frequency comb is shaped by classical processing circuits so that each frequency tooth carries a designated phase to address its corresponding quantum spectral mode. The processed classical frequency comb interferes with the 3D cluster state at a 50:50 beamsplitter, whose outputs are frequency demultiplexed by wavelength-division multiplexers (WDMs). An array of detectors perform balanced measurements. The measurement outcome is processed by a classical algorithm that determines the basis settings for homodyne measurements on the next batch of temporal modes.


\section{Conclusion} \label{SEC:conc}
We have proposed and analyzed a scalable and experimentally viable architecture for generating time-frequency-multiplexed 3D CV cluster states and utilizing them for large-scale quantum computing in integrated photonic circuits. Our architecture inherits the compactness of previous bulk-optical approaches, and thus, only requires  a handful of squeezed light sources, multimode interference couplers, delay lines, and Mach-Zehnder interferometers. The squeezed light is produced via a $\chi^{(3)}$ nonlinearity enhanced by MRs. This platform offers two key advantages: 1) only constant length delay lines are required to grow the resource state in the time direction; 2) and we can employ frequency multiplexing to extend the state in the frequency domain, and it is possible to address a large number (>2000) spectral modes via a frequency-comb soliton local oscillator. This claim is backed up by our numerical analysis, which found compatible physical device parameters to generate such a 3D cluster state. The extension of previous schemes~\cite{Menicucci11, Wang14, Alexander16, Alexander18, Larsen19} to the 3D case is significant as it opens the door topological error correction for Bosonic qubits~\cite{Vuillot19,Fukui18}.

Our architecture has the added benefit that it can be reprogrammed to generate cluster states of dimension less than three. This makes it compatible with previously studied protocols for lower dimensional cluster states~\cite{Alexander16a, Alexander17b, Menicucci18}.

\begin{acknowledgments}
B.-H.~W.~and Z.~Z.~are supported by the Office of Naval Research Award No.~N00014-19-1-2190 and the National Science Foundation Award No.~{ECCS-1920742}. R.~N.~A.~is supported by National Science Foundation Award No.~{PHY-1630114}. S.~L.~is supported by the Arizona Board of Regents Innovation Funds. Z.~Z. thanks the University of Arizona for providing startup funds.
\end{acknowledgments}

\appendix

\section{Graphical notation}\label{sec:graphnot}
In this article we describe how to generate various multimode Gaussian states. It will be convenient to use a graphical representation of each state~\cite{Menicucci11a}. This allows each Gaussian pure state to be represented up to displacements and overall phase by a complex weighted adjacency matrix
\begin{align}
    \mathbf{Z}\equiv \mathbf{V} + i\mathbf{U}.
\end{align}
We define $\hat{\mathbf{x}} \equiv (\hat{q}_{1},\dots, \hat{q}_{n}, \hat{p}_{1}, \dots, \hat{p}_{n})^{\text{T}}$, where $\hat{a}=(\hat{q} + i\hat{p})/\sqrt{2}$. 

For CV cluster states, the visual representation of the corresponding graph gives a direct indication of the correlations between various modes, and can in geneneral be related to the correlation matrix of the state's Wigner function 
\begin{align}
    \Sigma_{jk} = \frac{1}{2}\left\langle\left\{\hat{x}_{j}, \hat{x}_{k}\right\}\right\rangle
\end{align}
via the equation~\cite{Menicucci11a}
\begin{align}
    \bm{\Sigma} = \frac{1}{2}
    \begin{pmatrix} 
    \mathbf{U}^{-1} & \mathbf{U}^{-1}\mathbf{V} \\
    \mathbf{V}^{-1} \mathbf{U} & \mathbf{U} + \mathbf{V} \mathbf{U}^{-1} \mathbf{V}
    \end{pmatrix}.
\end{align}

This graphical description can be made even simpler when describing states whose non-zero graph edges all have same magnitude and differ only by a sign. 
We will not show self loops, which will all take value $i\,\text{cosh}2r$, where $r$ is a parameter that describes the overall squeezing used to produce the state. Blue and yellow edges are all weight $\mp i C \,\text{sinh}2r$, respectively, where $C$ is a real rescaling parameter specified in the relevant figure captions throughout this Article. The states represented by these graphs are not CV cluster states. However, all states described in this paper have the property that they are made from two-mode squeezed states and beamsplitters that do not mix the position and momentum quadratures. As a result, they can be converted to CV cluster states with self-loop weights $i\,\text{sech}2r$ and edge weights $\pm C\,\text{tanh}2r$, respectively, by applying $\pi/4$-phase delays prior to each measurement, which can be incorporated into the local oscillator(for a proof, see Theorem 1 in Ref.~\cite{Alexander18}). Due to this fact, we will treat each Gaussian state as if it were such a CV cluster state. These conventions are the same as those used and described in Refs.~\cite{Alexander16, Alexander16a, Alexander18}. 

\begin{figure}
    \centering
    \includegraphics[width=0.95\linewidth]{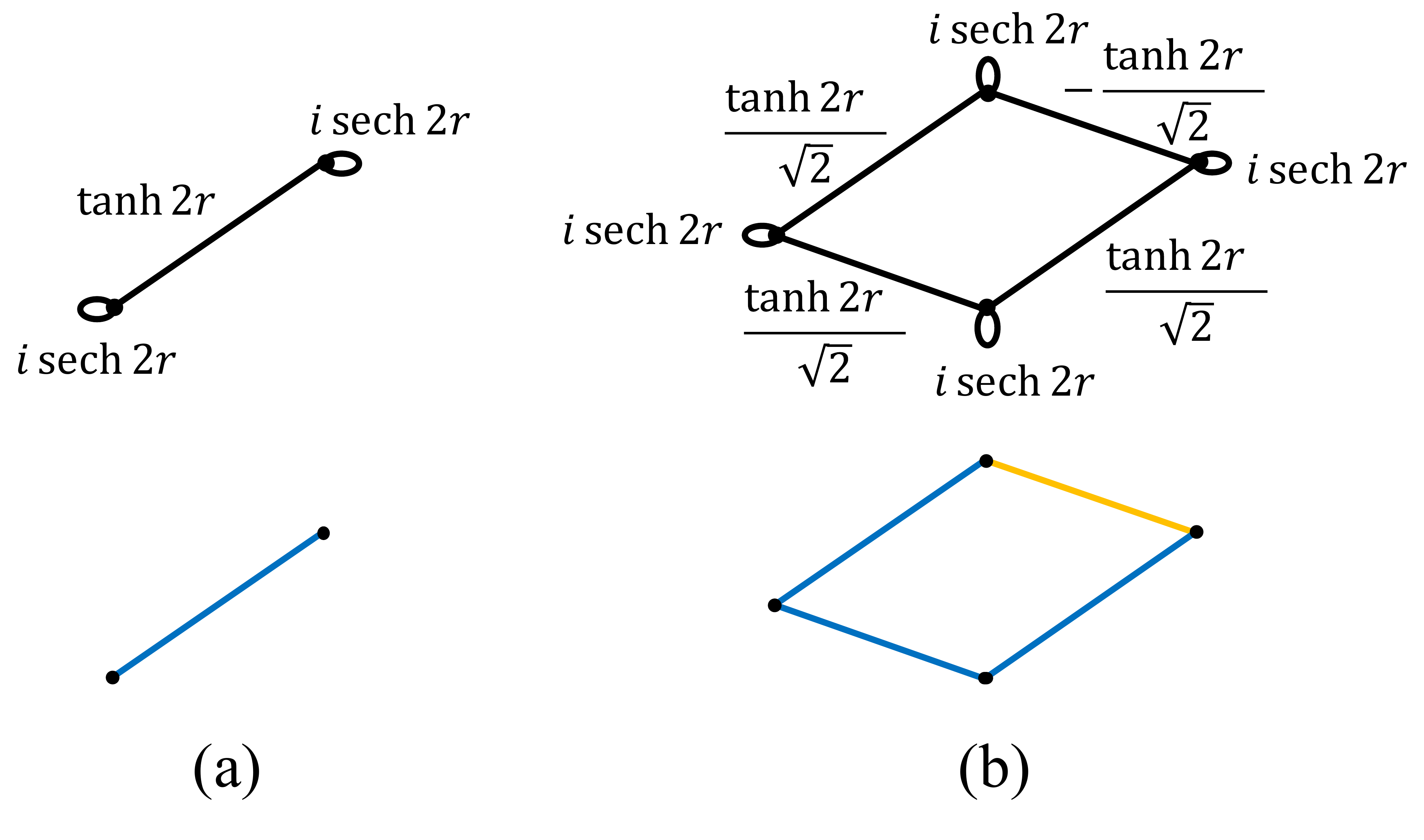}
    \caption{On top we show the desription of the Gaussian pure state using the full graphical calculus. On the bottom we show the simplified notation described in the main text. (a) Two-mode continuous-variable cluster state. The rescaling parameter is $C=1$. (b) Four mode continuous-variable cluster state. The rescaling parameter is $C=1/\sqrt{2}$. }
    \label{FIG:graphsimp}
\end{figure}

Some examples of the simplified graphical notation are given in Fig.~\ref{FIG:graphsimp}. 

The evolution of Gaussian pure states under Gaussian unitaries can be incorporated into this formalism as a graphical update rule~\cite{Menicucci11a}. One transformation that is particularly useful for understanding the construction of CV cluster states is the \emph{beamsplitter rule}:
\begin{align}
    \includegraphics[width=0.4\linewidth]{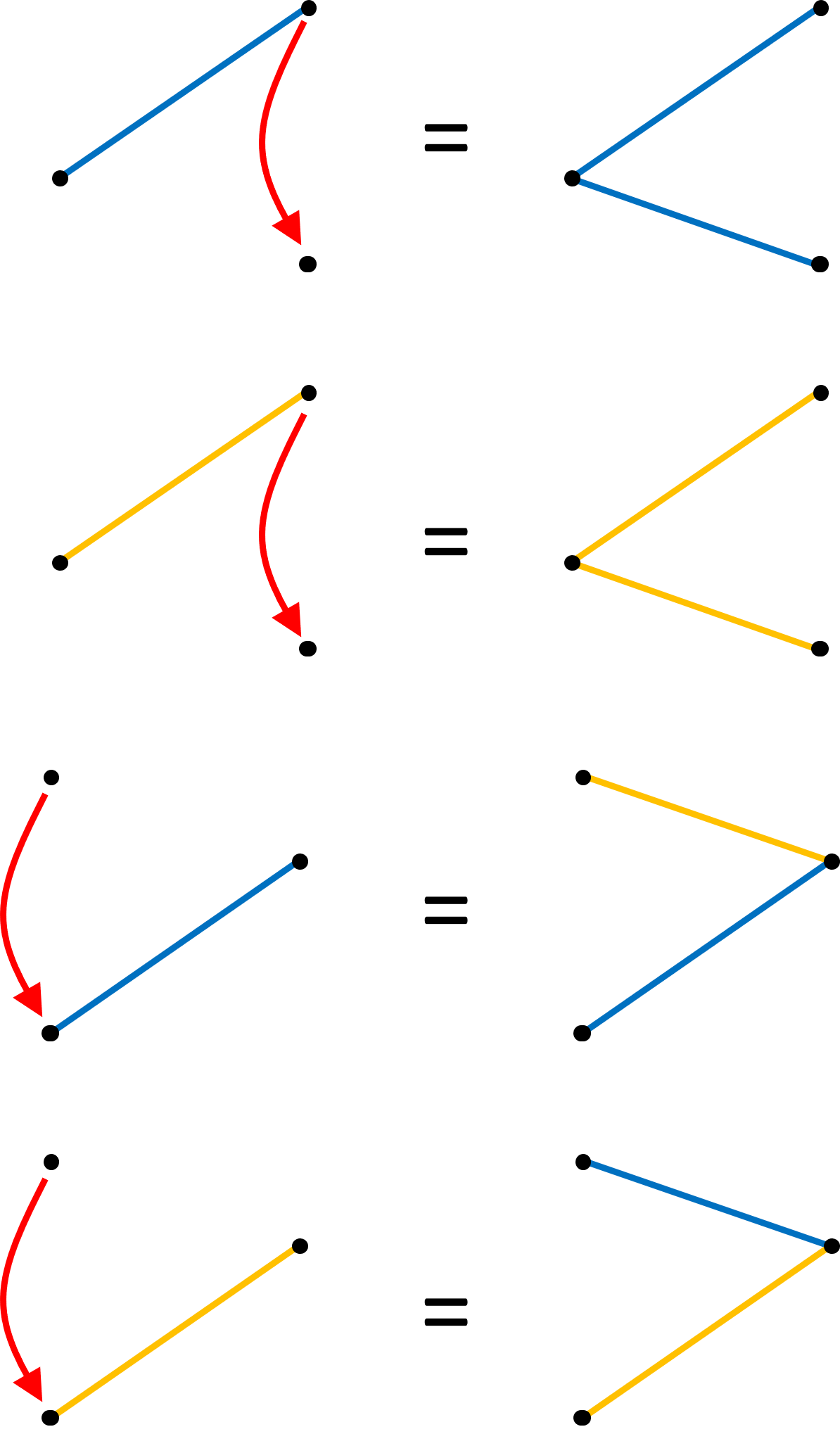}
\end{align}
, where the arrow points from mode $j$ to mode $k$, indicating the application of a 50:50 beamsplitter $\hat{B}_{jk}$. Let $\mathcal{C}$ by the rescaling parameter for the left hand side. Then the rescaling parameter for the right hand side is $\mathcal{C}/\sqrt{2}$~\cite{Menicucci11}. Note that the beamsplitter ``copies'' each link, up to a change in sign.

\section{\label{SEC:Disp}Dispersion parameters $\zeta_{n}$}
The dispersion effect shifts the cavity resonant frequency nonuniformly. Given the dispersion parameters $\beta_{1}$, $\beta_{2}$, $\beta_{3}\cdots$, here we seek to solve for the coefficients $\zeta_{2}$, $\zeta_{3}\cdots$ via Eq.~(\ref{EQ:cavityfreq}). 

Intracavity fields are constrained by periodic boundary conditions, $\beta(\omega_{l})L=2\pi\times l$, $\forall l\in\field{Z}$, where $L$ is the circumference of MR cavity, and we have
\begin{equation}
\label{EQ:Zeta}    
    \sum_{n=0}^{\infty}\frac{\beta_{n}}{n!}(l\Delta\omega+\sum_{m=2}^{\infty}\frac{\zeta_{m}}{m!}l^{m})^{n}L=2\pi l.
\end{equation}
We match the coefficients, $l$, $l^{2}$, $l^{3}\cdots$ for both sides of Eq.~(\ref{EQ:Zeta}) and derive $\zeta_{2}$ and $\zeta_{3}$,
\begin{equation}
\label{EQ:Zeta2} 
    \begin{aligned}
    \zeta_{2}&=-\frac{4\pi^2\beta_{2}}{L^2\beta_{1}^3},\\
    \zeta_{3}&=\frac{8\pi^3\left(3\beta_{2}^2-\beta_{1}\beta_{3}\right)}{L^3\beta_{1}^5}.
    \end{aligned}
\end{equation}
From Eq.~(\ref{EQ:Zeta2}), we calculate the dispersion-induced resonant-frequency shifts by determining $\zeta_{2}$ and $\zeta_{3}$ or even higher order terms.


\nocite{*}

\begin{thebibliography}{99}












\bibitem{Shor94}
P. W. Shor, {\em Scheme for reducing decoherence in quantum computer memory}, Phys. Rev. A, {\bf 52}, R2493(R) (1995).

\bibitem{Wiebe12}
N. Wiebe, D. Braun, and S. Lloyd, {\em Quantum Algorithm for Data Fitting}, Phys. Rev. Lett. {\bf 109}, 050505 (2012).

\bibitem{Djidjev18}
H. N. Djidjev, G. Chapuis, G. Hahn, and G. Rizk, {\em Efficient Combinatorial Optimization Using Quantum Annealing}, arXiv:1801.08653.

\bibitem{Aaronson10}
S. Aaronson and A. Arkhipov, {\em The computational complexity of linear optics}, arXiv:1011.3245.

\bibitem{Ganzhorn19}
M. Ganzhorn {\em et al.}, {\em Gate-Efficient Simulation of Molecular Eigenstates on a Quantum Computer}, Phys. Rev. Appl. {\bf 11}, 044092 (2019).

\bibitem{Monroe95}
C. Monroe, D. M. Meekhof, B. E. King, W. M. Itano, and D. J. Wineland, {\em Demonstration of a Fundamental Quantum Logic Gate}, Phys. Rev. Lett. {\bf 75}, 4714 (1995).

\bibitem{Wallraff04}
A. Wallraff, D. I. Schuster, A. Blais, L. Frunzio, R.-S. Huang, J. Majer, S. Kumar, S. M. Girvin, and R. J. Schoelkopf, {\em Strong coupling of a single photon to a superconducting qubit using circuit quantum electrodynamics}, Nature {\bf 431}, 162 (2004).

\bibitem{Zhou17}
B. B. Zhou, P. C. Jerger, V. O. Shkolnikov, F. J. Heremans, G. Burkard, and D. D. Awschalom, {\em Holonomic quantum control by coherent optical excitation in diamond}, Phys. Rev. Lett. {\bf 119}, 140503 (2017).

\bibitem{Loss98}
D. Loss and D. P. DiVincenzo, {\em Quantum computing with quantum dots}, Phys. Rev. A {\bf 57}, 120 (1998).

\bibitem{Raussendorf01}
R. Raussendorf and H. J. Briegel, {\em A One-Way Quantum Computer}, Phys. Rev. Lett. {\bf 86}, 5188 (2001). 

\bibitem{Browne05}
D. E. Browne and T. Rudolph, {\em Resource-Efficient Linear Optical Quantum computing}, Phys. Rev. Lett. {\bf 95}, 010501 (2005).

\bibitem{Briegel01}
H. J. Briegel and R. Raussendorf, {\em Persistent Entanglement in Arrays of Interacting Particles}, Phys. Rev. Lett. {\bf 86}, 910 (2001).

\bibitem{Nielsen04}
M. A. Nielsen, {\em Optical Quantum computing Using Cluster States}, Phys. Rev. Lett. {\bf 93}, 040503 (2004). 


\bibitem{Joo07}
J. Joo, P. L. Knight, J. L. O'Brien, and T. Rudolph, {\em One-way quantum computing with four-dimensional photonic qudits}, Phys. Rev. A {\bf 76}, 052326 (2007).

\bibitem{Walther05}
P. Walther, K. J. Resch, T. Rudolph, E. Schenck, H. Weinfurter, V. Vedral, M. Aspelmeyer, and A. Zeilinger, {\em Experimental one-way quantum computing}, Nature {\bf 434}, 169 (2005).

\bibitem{Kiesel05}
N. Kiesel, C. Schmid, U. Weber, G. T\'oth, O. G\"uhne, R. Ursin, and H. Weinfurter, {\em Experimental Analysis of a Four-Qubit Photon Cluster State}, Phys. Rev. Lett. {\bf 95}, 210502 (2005).

\bibitem{Biggerstaff09}
D. N. Biggerstaff, R. Kaltenbaek, D. R. Hamel, G. Weihs, T. Rudolph, and K. J. Resch, {\em Cluster-State Quantum Computing Enhanced by High-Fidelity Generalized Measurements}, Phys. Rev. Lett. {\bf 103}, 240504 (2009). 

\bibitem{Vallone07}
G. Vallone, E. Pomarico, P. Mataloni, F. De Martini, and V. Berardi, {\em Realization and Characterization of a Two-Photon Four-Qubit Linear Cluster State}, Phys. Rev. Lett. {\bf 98}, 180502 (2007). 


\bibitem{Russo19}
A. Russo, E. Barnes, and S. E. Economou, {\em Generation of arbitrary all-photonic graph states from quantum emitters}, New J. Phys. {\bf 21} 055002 (2019).

\bibitem{Buterakos17}
D. Buterakos, E. Barnes, and S. E. Economou, {\em Deterministic Generation of All-Photonic Quantum Repeaters from Solid-State Emitters}, Phys. Rev. X {\bf 7}, 041023 (2017).




\bibitem{Braunstein00}
S. L. Braunstein and H. J. Kimble, {\em Dense coding for continuous variables}, Phys. Rev. A {\bf 61}, 042302 (2000).

\bibitem{Braunstein98}
S. L. Braunstein and H. J. Kimble, {\em Teleportation of Continuous Quantum Variables}, Phys. Rev. Lett. {\bf 80}, 869 (1998).

\bibitem{Ralph99}
T. C. Ralph, {\em Continuous variable quantum cryptography}, Phys. Rev. A {\bf 61}, 010303(R) (1999).

\bibitem{Menicucci06}
N. C. Menicucci, P. v. Loock, M. Gu, C. Weedbrook, T. C. Ralph, and M. A. Nielsen, {\em Universal Quantum computing with Continuous-Variable Cluster States}, Phys. Rev. Lett. {\bf 97}, 110501 (2006).

\bibitem{Chen14}
M. Chen, N. C. Menicucci, and O. Pfister, {\em Experimental Realization of Multipartite Entanglement of 60 Modes of a Quantum Optical Frequency Comb}, Phys. Rev. Lett. {\bf 112}, 120505 (2014). 

\bibitem{Menicucci08}
N. C. Menicucci, S. T. Flammia, and O. Pfister, {\em One-Way Quantum Computing in the Optical Frequency Comb}, Phys. Rev. Lett. {\bf 101}, 130501 (2008).

\bibitem{Menicucci11}
N. C. Menicucci, {\em Temporal-mode continuous-variable cluster states using linear optics}, Phys. Rev. A {\bf 83}, 062314 (2011).

\bibitem{Yokoyama13}
S. Yokoyama, R. Ukai, S. C. Armstrong, C. Sornphiphatphong, T. Kaji, S. Suzuki, J. Yoshikawa, H. Yonezawa, N. C. Menicucci, and Akira Furusawa, {\em Ultra-large-scale continuous-variable cluster states multiplexed in the time domain}, Nat. Photonics {\bf 7}, 982 (2013). 

\bibitem{Alexander18}
R. N. Alexander, S. Yokoyama, A. Furusawa, and N. C. Menicucci, {\em Universal quantum computing with temporal-mode bilayer square lattices}, Phys. Rev. A {\bf 97}, 032302 (2018).

\bibitem{Sabapathy18}
D. Su, K. K. Sabapathy, C. R. Myers, H. Qi, C. Weedbrook, and Kamil Br\'adler, {\em Implementing quantum algorithms on temporal photonic cluster states}, Phys. Rev. A {\bf 98}, 032316 (2018).

\bibitem{Yoshikawa16}
J. Yoshikawa, S. Yokoyama, T. Kaji, C. Sornphiphatphong Y. Shiozawa, K. Makino, and A. Furusawa, {\em Generation of one-millionmode continuous-variable cluster state by unlimited time-domain multiplexing}, APL Photonics {\bf 1}, 060801 (2016).

\bibitem{Asavanant19}
W. Asavanant {\em et al.}, {\em Time-Domain Multiplexed 2-Dimensional Cluster State: Universal Quantum Computing Platform}, arXiv:1903.03918.

\bibitem{Larsen19}
M. V. Larsen, X. Guo, C. R. Breum, J. S. Neergaard-Nielsen, and U. L. Andersen, {\em Deterministic generation of a two-dimensional cluster state for universal quantum computing}, arXiv:1906.08709.

\bibitem{Alexander16}
R. N. Alexander, P. Wang, N. Sridhar, M. Chen, O. Pfister, and N. C. Menicucci, {\em One-way quantum computing with arbitrarily large time-frequency continuous-variable cluster states from a single optical parametric oscillator}, Phys. Rev. A {\bf 94}, 032327 (2016).

\bibitem{Humphreys14}
P. C. Humphreys, W. S. Kolthammer, J. Nunn, M. Barbieri, A. Datta, and I. A. Walmsley, {\em Continuous-Variable Quantum Computing in Optical Time-Frequency Modes Using Quantum Memories}, Phys. Rev. Lett. {\bf 113}, 130502 (2014).

\bibitem{Gu09}
M. Gu, C. Weedbrook, N. C. Menicucci, T. C. Ralph, and P. v. Loock, {\em Quantum computing with continuous-variable clusters}, Phys. Rev. A {\bf79}, 062318 (2009). 

\bibitem{Alexander14}
R. N. Alexander, S. C. Armstrong, R. Ukai, and N. C. Menicucci, {\em Noise analysis of single-mode Gaussian operations using continuous-variable cluster states}, Phys. Rev. A {\bf 90}, 062324 (2014).

\bibitem{Menicucci14}
N. C. Menicucci, {\em Fault-Tolerant Measurement-Based Quantum Computing with Continuous-Variable Cluster States} Phys. Rev. Lett. {\bf 112}, 120504 (2014).

\bibitem{Gottesman01}
D. Gottesman, A. Kitaev, and J. Preskill, {\em Encoding a qubit in an oscillator}, Phys. Rev. A, {\bf 64}, 012310 (2001).

\bibitem{Fukui17}
K. Fukui, A. Tomita, and A. Okamoto, {\em Analog Quantum Error Correction with Encoding a Qubit into an Oscillator}, Phys. Rev. Lett., {\bf 119}, 180507 (2017). 

\bibitem{Vuillot19}
C. Vuillot, H. Asasi, Y. Wang, L. P. Pryadko, and B. M. Terhal, {\em Quantum error correction with the toric Gottesman-Kitaev-Preskill code}, Phys. Rev. A, {\bf 99}, 032344 (2019).

\bibitem{Fukui19}
K. Fukui, {\em High-threshold fault-tolerant quantum computation with the GKP qubit and realistically noisy devices}, arXiv:1906.09767 (2019).

\bibitem{Noh19}
K. Noh, C. Chamberland {\em Fault-tolerant bosonic quantum error correction with the surface-GKP code}, arXiv:1908.03579 (2019).

\bibitem{Fluhmann19}
C. Fl\"uhmann, T. L. Nguyen, M. Marinelli, V. Negnevitsky, K. Mehta, and J. P. Home, {\em Encoding a qubit in a trapped-ion mechanical oscillator}, Nature {\bf566}, 513 (2019).

\bibitem{Touzard19}
P. Campagne-Ibarcq {\em et al.}, {\em A stabilized logical quantum bit encoded in grid states of a superconducting cavity}, arXiv:1907.12487.


\bibitem{Lamont13}
M. R. E. Lamont, Y. Okawachi, and A. L. Gaeta, {\em Route to stabilized ultrabroadband microresonator-based frequency combs}, Opt. Lett. {\bf 38}, 3478 (2013).

\bibitem{Coen13}
S. Coen, H. G. Randle, T. Sylvestre, and M. Erkintalo, {\em Modeling of octave-spanning Kerr frequency combs using a generalized mean-field Lugiato-Lefever model}, Opt. Lett. {\bf 38}, 37 (2013).

\bibitem{Zhang13}
Y. Zhang, S. Yang, A. E.-J. Lim, G.-Q. Lo, C. Galland, T. Baehr-Jones, and M. Hochberg, {\em A compact and low loss Y-junction for submicron silicon waveguide}, Opt. Express {\bf 21}, 1310 (2013).

\bibitem{Glockl04}
O. Gl\"{o}ckl, U. L. Andersen, S. Lorenz, Ch. Silberhorn, N. Korolkova, and G. Leuchs, {\em Sub-shot-noise phase quadrature measurement of intense light beams}, Opt. Lett. {\bf 29}, 1936 (2004).

\bibitem{Huntington05}
E. H. Huntington, G. N. Milford, C. Robilliard, T. C. Ralph, O. Gl\"ockl, U. L. Andersen, S. Lorenz, and G. Leuchs, {\em Demonstration of the spatial separation of the entangled quantum sidebands of an optical field}, Phys. Rev. A {\bf 71}, 041802(R) (2005).

\bibitem{Elshaari16}
A. W. Elshaari, I. E. Zadeh, K. D. J\"{o}ns, and V. Zwiller, {\em Thermo-Optic Characterization of Silicon Nitride Resonators for Cryogenic Photonic Circuits}, IEEE Photonics J. {\bf 8}, 2701009 (2016).

\bibitem{Xue16}
X. Xue, Y. Xuan, C. Wang, P.-H. Wang, Y. Liu, B. Niu, D. E. Leaird, M. Qi, and A. M. Weiner, {\em Thermal tuning of Kerr frequency combs in silicon nitride microring resonators}, Opt. Express {\bf 24}, 687 (2016).

\bibitem{Wang14}
P. Wang, M. Chen, N. C. Menicucci, and O. Pfister, {\em Weaving quantum optical frequency combs into continuous-variable hypercubic cluster states}, Phys. Rev. A {\bf 90}, 032325 (2014).

\bibitem{Loock03}
P. v. Loock and A. Furusawa, {\em Detecting genuine multipartite continuous-variable entanglement}, Phys. Rev. Lett. {\bf 67}, 052315 (2003). 

\bibitem{Lipson05}
M. Lipson, {\em Guiding, Modulating, and Emitting Light on Silicon-Challenges and Opportunities}, IEEE J. Lightw. Technol. {\bf 23}, 4222 (2005).

\bibitem{Clemmen09}
S. Clemmen, K. P. Huy, W. Bogaerts, R. G. Baets, P. Emplit, and S. Massar, {\em Continuous wave photon pair generation in silicon-on-insulator waveguides and ring resonators}, Opt. Express {\bf 17}, 16558 (2009).

\bibitem{Sharping06}
J. E. Sharping, K. F. Lee, M. A. Foster, A. C. Turner, B. S. Schmidt, M. Lipson, A. L. Gaeta, and P. Kumar, {\em Generation of correlated photons in nanoscale silicon waveguides}, Opt. Express {\bf 14}, 12388 (2006).

\bibitem{Davanco12}
M. Davanco, J. R. Ong, A. B. Shehata, A. Tosi, I. Agha, S. Assefa, F. Xia, W. M. J. Green, S. Mookherjea, and K. Srinivasan, {\em Telecommunications-band heralded single photons from a silicon nanophotonic chip}, Appl. Phys. Lett. {\bf 100}, 261104 (2012).

\bibitem{Takesue08}
H. Takesue, H. Fukuda, T. Tsuchizawa, T. Watanabe, K. Yamada, Y. Tokura, and S. Itabashi, {\em Generation of polarization entangled photon pairs using silicon wire waveguide}, Opt. Express {\bf 16}, 5721 (2008).

\bibitem{Najafi15}
F. Najafi {\em et al.}, {\em On-chip detection of non-classical light by scalable integration of single-photon detectors}, Nat. Commun. {\bf 6}, 5873 (2015).

\bibitem{O'Brien09}
J. L. O'Brien, A. Furusawa, and J. Vu$\check{\text{c}}$kovi$\acute{\text{c}}$, {\em Photonic quantum technologies}, Nat. Photonics {\bf 3}, 687 (2009).

\bibitem{Wang18}
J. Wang {\em et al.}, {\em Multidimensional quantum entanglement with large-scale integrated optics}, Science {\bf 360}, 285 (2018).

\bibitem{Kuyken15}
B. Kuyken {\em et al.}, {\em An octave-spanning mid-infrared frequency comb generated in a silicon nanophotonic wire waveguide}, Nat. Commun. {\bf 6}, 6310 (2015).

\bibitem{Lenzini18}
F. Lenzini {\em et al.}, {\em Integrated photonic platform for quantum information with continuous variables}, Sci. Adv. {\bf 4}, eaat9331 (2018).

\bibitem{Mondain19}
F. Mondain, T. Lunghi, A. Zavatta, E. Gouzien, F. Doutre, M. De Micheli, S. Tanzilli, and V. D'Auria, {\em Chip-based squeezing at a telecom wavelength}, Photon. Res. {\bf 7}, A36 (2019).

\bibitem{Moss13}
D. J. Moss, R. Morandotti, A. L. Gaeta, and M. Lipson, {\em New CMOS-compatible platforms based on silicon nitride and Hydex for nonlinear optics}, Nat. Photonics {\bf7}, 597 (2013).

\bibitem{Dutt15}
A. Dutt, K. Luke, S. Manipatruni, A. L. Gaeta, P. Nussenzveig, and M. Lipson, {\em On-Chip Optical Squeezing}, Phys. Rev. Appl. {\bf 3}, 044005 (2015).

\bibitem{Dutt16}
A. Dutt, S. Miller, K. Luke, J. Cardenas, A. L. Gaeta, P. Nussenzveig, and M. Lipson, {\em Tunable squeezing using coupled ring resonators on a silicon nitride chip}, Opt. Lett. {\bf 41}, 223 (2016).

\bibitem{Ramelow15}
S. Ramelow, A. Farsi, S. Clemmen, D. Orquiza, K. Luke, M. Lipson, and A. L. Gaeta, {\em Silicon-Nitride Platform for Narrowband Entangled Photon Generation}, arXiv:1508.04358.

\bibitem{Vaidya19}
V. D. Vaidya {\em et al.}, {\em Broadband quadrature-squeezed vacuum and nonclassical photon number correlations from a nanophotonic device}, arXiv:1904.07833.

\bibitem{Hoff15}
U. B. Hoff, B. M. Nielsen, and U. L. Andersen, {\em Integrated source of broadband quadrature squeezed light}, Opt. Express {\bf 23}, 12013 (2015).

\bibitem{Kippenberg18}
T. J. Kippenberg, A. L. Gaeta, M. Lipson, and M. L. Gorodetsky, {\em Dissipative Kerr solitons in optical microresonators}, Science {\bf 361}, eaan8083 (2018). 

\bibitem{Bao17}
C. Bao, H. Taheri, L. Zhang, A. Matsko, Y. Yan, P. Liao, L. Maleki, and A. E. Willner, {\em High-order dispersion in Kerr comb oscillators}, J. Opt. Soc. Am. B {\bf 34}, 715 (2017).

\bibitem{Hansson14}
T. Hansson, D. Modotto, and S. Wabnitz, {\em On the numerical simulation of Kerr frequency combs using coupled mode equations}, Opt. Commun. {\bf 312}, 134 (2014).

\bibitem{Chembo13}
Y. K. Chembo and C. R. Menyuk, {\em Spatiotemporal Lugiato-Lefever formalism for Kerr-comb generation in whispering-gallery-mode resonators}, Phys. Rev. A {\bf 87}, 053852 (2013). 

\bibitem{Pfeiffer17}
M. H. Pfeiffer, C. Herkommer, J. Liu, H. Guo, M. Karpov, E. Lucas, M. Zervas, and T. J. Kippenberg, {\em Octave-spanning dissipative Kerr soliton frequency combs in Si$_{3}$N$_{4}$ microresonators}, Optica {\bf 4}, 684 (2017).

\bibitem{Bauters11}
J. F. Bauters, M. J. R. Heck, D. John, D. Dai, M.-C. Tien, J. S. Barton, A. Leinse, R. G. Heideman, D. J. Blumenthal, and J. E. Bowers, {\em Ultra-low-loss high-aspect-ratio Si$_3$N$_4$ waveguides}, Opt. Express {\bf 19}, 3163 (2011).

\bibitem{Chembo16}
Y. K. Chembo, {\em Quantum dynamics of Kerr optical frequency combs below and above threshold: Spontaneous four-wave mixing, entanglement, and squeezed states of light}, Phys. Rev. A {\bf 93}, 033820 (2016). 

\bibitem{Chembo 10}
Y. K. Chembo and N. Yu, {\em Modal expansion approach to optical-frequency-comb generation with monolithic whispering-gallery-mode resonators}, Phys. Rev. A {\bf 82}, 033801 (2010). 

\bibitem{Chembo10}
Y. K. Chembo, D. V. Strekalov, and N. Yu, Spectrum, {\em Spectrum and Dynamics of Optical Frequency Combs Generated with Monolithic Whispering Gallery Mode Resonators}, Phys. Rev. Lett. {\bf 104}, 103902 (2010).

\bibitem{Herr13}
T. Herr, K. Hartinger, J. Riemensberger, C. Y. Wang, E. Gavartin, R. Holzwarth, M. L. Gorodetsky, and T. J. Kippenberg, {\em Universal formation dynamics and noise of Kerr-frequency combs in microresonators}, Nat. Photonics {\bf 6}, 480 (2012).

\bibitem{Hu17}
X. Hu, W. Wang, L. Wang, W. Zhang, Y. Wang, and W. Zhao, {\em Numerical simulation and temporal characterization of dual-pumped microring-resonator-based optical frequency combs}, Photon. Res. {\bf 5}, 207 (2017).

\bibitem{Xuan16}
Y. Xuan {\em et al.}, {\em High-Q silicon nitride microresonators exhibiting low-power frequency comb initiation}, Optica {\bf 3}, 1171 (2016).

\bibitem{Ji17}
X. Ji, F. A. S. Barbosa, S. P. Roberts, A. Dutt, J. Cardenas, Y. Okawachi, A. Bryant, A. L. Gaeta, and M. Lipson, {\em Ultra-low-loss on-chip resonators with sub-milliwatt parametric oscillation threshold}, Optica {\bf 4}, 619 (2017).

\bibitem{Zhang14}
L. Zhang, A. M. Agarwal, L. C. Kimerling, and J. Michel, {\em Nonlinear Group IV photonics based on silicon and germanium: from near-infrared to mid-infrared}, Nanophotonics {\bf 3}, 247 (2014).

\bibitem{Levy 10}
J. S. Levy, A. Gondarenko, M. A. Foster, A. C. Turner-Foster, A. L. Gaeta, and M. Lipson, {\em CMOS-compatible multiple-wavelength oscillator for on-chip optical interconnects}, Nat. Photonics {\bf 4}, 37 (2010).

\bibitem{Haus95}
H. A. Haus, {\em From classical to quantum noise}, J. Opt. Soc. Am. B {\bf 12}, 2019 (1995).

\bibitem{Pysher11}
See supplementary information of"{\em Parallel Generation of Quadripartite Cluster Entanglement in the Optical Frequency Comb}", M. Pysher, Y. Miwa, R. Shahrokhshahi, R. Bloomer, and O. Pfister, Phys. Rev. Lett. {\bf 107}, 030505 (2011). 



\bibitem{Menicucci07}
N. C. Menicucci, {\em Temporal-mode continuous-variable cluster states using linear optics}, Phys. Rev. A {\bf 83}, 062314 (2011).

\bibitem{Blumenthal18}
D. J. Blumenthal, R. Heideman, D. Geuzebroek, A. Leinse, and C. Roeloffzen, {\em Silicon Nitride in Silicon
Photonics}, in {\em Proceedings of the IEEE, 2018}, pp. 2209-2231.

\bibitem{Xu18}
X. Xu, J. Wu, S. T. Chu, B. E. Little, R. Morandotti, A. Mitchell, and D. J. Moss, {\em Emerging applications of integrated optical microcombs for analogue RF and microwave photonic signal processing}, arXiv:1808.04462.

\bibitem{Alexander16a}
R. N. Alexander and N. C. Menicucci, {\em Flexible quantum circuits using scalable continuous-variable cluster states}, Phys. Rev. A {\bf 93}, 062326 (2016).

\bibitem{Ukai10}
R. Ukai, J.-i. Yoshikawa, N. Iwata, P. van Loock, and A. Furusawa, {\em Universal linear Bogoliubov transformations through one-way quantum computing} Phys. Rev. A {\bf 81}, 032315 (2010).

\bibitem{Baragiola19}
B. Q. Baragiola, G. Pantaleoni, R. N. Alexander, A. Karanjai, and N. C. Menicucci, {\em All-Gaussian universality and fault tolerance with the Gottesman-Kitaev-Preskill code}, arXiv:1903.00012 (2019).

\bibitem{Fukui18}
K. Fukui, A. Tomita, A. Okamoto, and K. Fujii, {\em High-Threshold Fault-Tolerant Quantum Computation with Analog Quantum Error Correction}, Phys. Rev. X {\bf 8}, 021054 (2018).

\bibitem{Alexander17b}
R. N. Alexander, N. C. Gabay, P. P. Rohde, and N. C. Menicucci, {\em Measurement-based linear optics}, Phys. Rev. Lett. {\bf 118}, 110503 (2017).

\bibitem{Menicucci18}
N. C. Menicucci, B. Q. Baragiola, T. F. Demarie, and G. K. Brennen, {\em Anonymous broadcasting of classical information with a continuous-variable topological quantum code}, Phys. Rev. A {\bf 97}, 032345 (2018).

\bibitem{Menicucci11a}
N. C. Menicucci, S. T. Flammia, and P. v. Loock, {\em Graphical calculus for Gaussian pure states}, Phys. Rev. A {\bf83}, 042335 (2011).





























\end{thebibliography}
\end{document}